\documentclass[12pt]{article}

\usepackage{amsmath,amsthm, amsfonts, amssymb, amsxtra,amsopn}
\usepackage{pgfplots}
\usepgfplotslibrary{colormaps}
\pgfplotsset{compat=1.15}
\usepackage{pgfplotstable}
\usetikzlibrary{pgfplots.statistics} 
\usepackage{filecontents}
\usepackage{graphicx}
\usepackage{multirow}
\usepackage{tabu}
\usepackage{algorithmic}
\usepackage{longtable}
\usepackage{booktabs}
\usepackage{listings}
\usepackage{cmap}
\usepackage{colortbl}
\usepackage{adjustbox}
\usepackage{epsfig}

\usepackage{subfigure}

\PassOptionsToPackage{hyphens}{url}

\usepackage{hyperref}
\hypersetup{colorlinks=true,linkcolor=black,citecolor=black,urlcolor=blue,filecolor=black}
\hypersetup{pdfpagemode=UseNone,pdfstartview=}

\usepackage[tableposition=top]{caption}
\usepackage{enumitem}
\setlist[itemize]{noitemsep, topsep=0pt}

\advance\oddsidemargin by -0.45in
\advance\textwidth by 0.9in

\advance\topmargin by -0.3in
\advance\textheight by 0.6in

\long\def\symbolfootnotetext[#1]#2{\begingroup%
\def\thefootnote{\fnsymbol{footnote}}\footnotetext[#1]{#2}\endgroup}

\DeclareMathOperator{\AUC}{AUC}

\def\kNN{{$k$-NN}}

\pgfkeys{
    /pgf/number format/fixed zerofill=true }
    
\pgfplotstableset{
    color cells/.code={%
        \pgfqkeys{/color cells}{#1}%
        \pgfkeysalso{%
            postproc cell content/.code={%
                \begingroup
                \pgfkeysgetvalue{/pgfplots/table/@preprocessed cell content}\value
\ifx\value\empty
\endgroup
\else
                \pgfmathfloatparsenumber{\value}%
                \pgfmathfloattofixed{\pgfmathresult}%
                \let\value=\pgfmathresult
                \pgfplotscolormapaccess
                    [\pgfkeysvalueof{/color cells/min}:\pgfkeysvalueof{/color cells/max}]%
                    {\value}%
                    {\pgfkeysvalueof{/pgfplots/colormap name}}%
                \pgfkeysgetvalue{/pgfplots/table/@cell content}\typesetvalue
                \pgfkeysgetvalue{/color cells/textcolor}\textcolorvalue
                \toks0=\expandafter{\typesetvalue}%
                \xdef\temp{%
                    \noexpand\pgfkeysalso{%
                        @cell content={%
                            \noexpand\cellcolor[rgb]{\pgfmathresult}%
                            \noexpand\definecolor{mapped color}{rgb}{\pgfmathresult}%
                            \ifx\textcolorvalue\empty
                            \else
                                \noexpand\color{\textcolorvalue}%
                            \fi
                            \the\toks0 %
                        }%
                    }%
                }%
                \endgroup
                \temp
\fi
            }%
        }%
    }
}

\title{Multifamily Malware Models}

\author{Samanvitha Basole\footnotemark[1]\ \ \ 
Fabio Di Troia\footnotemark[1]\ \ \
Mark Stamp\footnotemark[1]\,\,\footnotemark[2]}

\begin{document}

\symbolfootnotetext[1]{Department of Computer Science, San Jose State University}
\symbolfootnotetext[2]{mark.stamp$@$sjsu.edu}

\maketitle

\abstract
When training a machine learning model, there is likely to be a tradeoff 
between accuracy and the diversity of the dataset. 
Previous research has shown that if we train a model to detect one specific 
malware family, we generally obtain stronger results
as compared to a case where we train a single model on multiple diverse families. 
However, during the detection phase, it would be more efficient to have a single model 
that can reliably detect multiple families, rather than having to score each sample against multiple models. 
In this research, we conduct experiments based on byte $n$-gram features
to quantify the relationship between the generality of the training dataset 
and the accuracy of the corresponding machine learning models, all within the 
context of the malware detection problem. We find that neighborhood-based
algorithms generalize surprisingly well, far outperforming the other
machine learning techniques considered.

\section{Introduction}\label{chap:intro}

Machine learning (ML) has been successfully applied to the malware detection 
problem; see, for example~\cite{rieck2011automatic,sahs2012machine}. Previous research has shown 
that ML approaches can perform better at detecting some types of malware than traditional 
signature-based techniques. For example,~\cite{sami2010malware} and~\cite{yerima2015high} 
achieve high detection rates with low false positive rates using machine learning techniques
on challenging malware datasets. 

In this paper, we consider 
the malware detection problem based on ``big data'' with the goal of 
carefully quantifying the effect of an increasingly generic dataset.  
Specifically, we examine the effectiveness of byte $n$-grams and the 
robustness of various machine learning techniques for detecting malware. 
The focus on $n$-grams is due, in part, to recent work~\cite{raff2018investigation}
that suggests such features are of little value in the malware detection domain.
The experimental results presented in this paper show that, to the contrary, 
$n$-grams can be a surprisingly strong feature, particularly when used in 
conjunction with neighborhood based machine learning algorithms.

Our experiments are based on a recently collected malware dataset that
is extremely large and diverse---being on the order of half a terabyte~\cite{bigdata}. 
As part of this work, we extract and analyze $n$-gram features
from this dataset. 

We consider byte bigram, 4-gram, and 6-gram features 
and employ a variety of machine learning techniques. 
Specifically, the machine learning techniques that we use are $k$-nearest neighbors (\kNN), 
support vector machines (SVM), random forests (RF), and 
multilayer perceptrons 
(MLP)~\cite{friedman2001elements,goodfellow2016deep,james2013introduction,stamp2017introduction}.
Both RF and \kNN\ can be viewed as neighborhood-based ML techniques, but
with substantially different neighborhood structure~\cite{LinJeon2006}. There is
also a close connection between SVMs and MLPs, as an MLP is somewhat analogous to an SVM where
the kernel function is learned, rather than being specified by the user~\cite{stamp2017introduction}.
Thus, we generally expect qualitatively similar results from~\kNN\ and RF, and the same
is true of the SVM and MLP pair.  We have more to say about these issues below.

To study the effect of an increasingly generic dataset, for each ML technique under
consideration, we first
train and test models for individual malware families. Then we consider models designed to 
detect pairs of families, triples of families, and so on, up to a single model for all~20 families
under consideration. In this way, we produce models that must deal with
progressively more generic datasets~\cite{bass18}. 

In practice, the fewer models that are required, the more efficient is
the detection process. However, intuitively, the more generic the dataset, the weaker
the models are likely to be. We carefully measure the accuracy of all of our models, which enables
us to quantify the tradeoff between the number of models
and the accuracy. This analysis also enables us to determine which of the 
machine learning techniques under consideration is best able to generalize
to the multifamily case. 

The remainder of this paper is organized as follows. Section~\ref{chap:background} discusses previous 
work related to this research problem, while Section~\ref{chap:methodology} outlines the methodology 
of our experiments. Section~\ref{chap:implementation} describes the malware families that we
have used, the inner workings of the machine learning techniques considered, and various
implementation details involved in this research. In Section~\ref{chap:results}, 
we present and analyze the results of our experiments. Finally, in Section~\ref{chap:conclusion}, 
we summarize this research and discuss possible avenues for future work.

\section{Related Work}\label{chap:background}

In this section, we first discuss the effectiveness of machine learning for classifying malware. 
We also discuss the motivation for---and previous work done---using byte $n$-gram features
in the malware domain. And, we provide a brief discussion of how this present research
augments previous related work.

Signature-based techniques have long been the workhorses 
in the field of malware detection. Such techniques rely on pattern matching. 
While fast and effective on traditional malware, signature-based techniques 
can be defeated by advanced forms of malware, as signatures 
cannot detect previously unseen malware. 

In recent years, machine learning (ML) has become a popular and highly successful method 
for malware detection. Machine learning offers the potential for vastly greater efficiency, 
since an individual model (or limited number of models) can be used to 
detect entire families or classes of malware. An additional advantage is that 
ML techniques can easily handle big datasets in the training phase.

Supervised machine learning involves training a model using labeled data and then 
validating the model predictions based on additional labeled data that was not part of the training set. 
In the malware domain, this process can be used to estimate the success of such models
when applied in the real world. 

Machine learning requires features or attributes that form the basis for classification. 
In the malware domain, examples of such features include opcodes, file headers, API calls, 
graph-based structures, and byte $n$-grams, among many others. 

An $n$-gram is a sequence of~$n$ consecutive features. Byte $n$-grams have been successfully 
and widely used as features for malware classification.
For example, in~\cite{santos2009n}, the authors use $n$-grams to detect unknown 
malware with a low false positive rate. They use the $k$-nearest neighbor (\kNN)
algorithm and experiment with different sizes of $n$-grams. They achieve their best 
results using 4-grams. 

In~\cite{liangboonprakong2013classification}, the authors achieve an accuracy 
of~96.64\%\ with~4-gram features and using support vector machines (SVM).
In addition to SVMs, these authors also experiment using decision trees and artificial neural networks
to classify malware into~10 families. Their dataset consists of~12,199 malware samples.

The research presented in~\cite{reddy2006n} employs a feature selection method 
using class-wise document frequency. They use the following steps to select 
relevant $n$-grams. 
\begin{itemize}
\item The $n$-grams are extracted and class-wise document frequencies are computed from 
both malware and benign samples.
\item The resulting $n$-grams are sorted in descending order.
\item The top $k$ most frequent $n$-grams are selected from each of the benign and malware sets.
\item The top $n$-grams from malware and benign are combined to form the working set of $n$-grams.
\end{itemize}
We follow a similar approach but since our goal is to model the malware samples, 
we do not consider the top $n$-grams from the benign samples. 

The authors in~\cite{raff2018investigation} claim that byte $n$-grams promote gross overfitting. 
They implement elastic-net regularized logistic regression~\cite{zou2005regularization}, 
and use a regularization path to examine the accuracy and other properties, as 
a function of the regularized parameter. 
The authors achieve low weighted accuracy based on their dataset. 
However, the authors of~\cite{raff2018investigation}
have used a large and, apparently, extremely diverse private dataset 
that was obtained from an undisclosed ``industry partner.'' In addition to 
making it impossible to independently replicate their results, it is not clear
that this dataset contains any well-defined malware families. Any dataset 
containing a vast number of sporadic malware samples is sure to 
blur the inherently blurry line between ``malware'' and ``benign.''
In contrast to~\cite{raff2018investigation}, we focus on clearly 
defined malware families, which offers the potential for extracting 
distinguishing features.

In~\cite{bass18}, the authors perform experiments using $n$-grams as features for 
different machine learning models. The classification techniques included SVM, 
a simple~$\chi^2$ test, \kNN, and random forest. Their feature extraction step involves selecting 
the~10 most-frequent $n$-grams from the benign set and the~10 most common $n$-grams from
the malware set, resulting in a feature vector of size at most~20. They find that a 
random forest classifier is more robust than other techniques and they determine the 
tradeoff between the generality of a model and the accuracy of its classification. 
Not surprisingly, the authors conclude that as the data becomes more generic, 
the accuracy declines.

Our research uses a similar approach as in~\cite{bass18}. That is, we use~$n$-grams 
to determine the tradeoff between the generality and the accuracy of a model. In contrast to~\cite{bass18}, 
we use~20 families instead of~8, a different feature selection method, and additional machine learning 
techniques, as discussed in more detail in Section~\ref{chap:implementation}. Our result lead to
a variety of insights that were not apparent from the smaller dataset used in~\cite{bass18}.
For example, we are able to compare the relative strengths and weaknesses of various 
machine learning techniques when applied to the specific malware detection problem
under consideration.

\section{Methodology}\label{chap:methodology}

In this section, we explain our feature extraction step and the classification experiments performed.
We also briefly discuss the machine learning pipeline used in each experiment. 

First, we collected a subset of~1000 benign samples and a malware dataset of~20 malware families, 
each family consisting of~1000 samples. Section~\ref{chap:implementation} includes some details 
on this process. Once this labeled dataset was available, $n$-gram features were extracted. Below, we 
discuss the process for bigram features---an analogous process is used in our 4-gram
and 6-gram experiments. 

\subsection{Feature Extraction}

The top~100 byte bigrams from each malware sample are stored as a 
dictionary consisting of the bigram and its frequency. 
We perform the same procedure for all benign samples, 
but since feature selection is based on malware families,
we store the top~500 bigrams and their frequencies, instead of the top~100. 
This approach enables us to efficiently determine bigram features for any combination
of families. 

\subsection{Classification Experiments}

We conduct several experiments to determine the effect on accuracy of increasingly
generic data. Our first experiment uses five-fold cross-validation to perform a binary 
classification of~1000 malware samples from one specific family versus our~1000 
representative benign samples. We perform such binary classification using each of
the~20 different malware families in our dataset. This can be viewed as a base case, against
which we can compare increasingly generic models.

At the second step (or level), we combine two different malware families to form the 
malware class, and then classify samples as benign or malware; at the third level,
we combine three different families, and so on.  
We consider a total of~20 families, and hence there are~20 levels.
At level~20, we combine all of the~20 malware families under consideration
to form the malware class, which results in our most general malware dataset. 

\subsection{Machine Learning Pipeline}

\begin{figure}[!htb]
\centering
\includegraphics[width=0.85\textwidth]{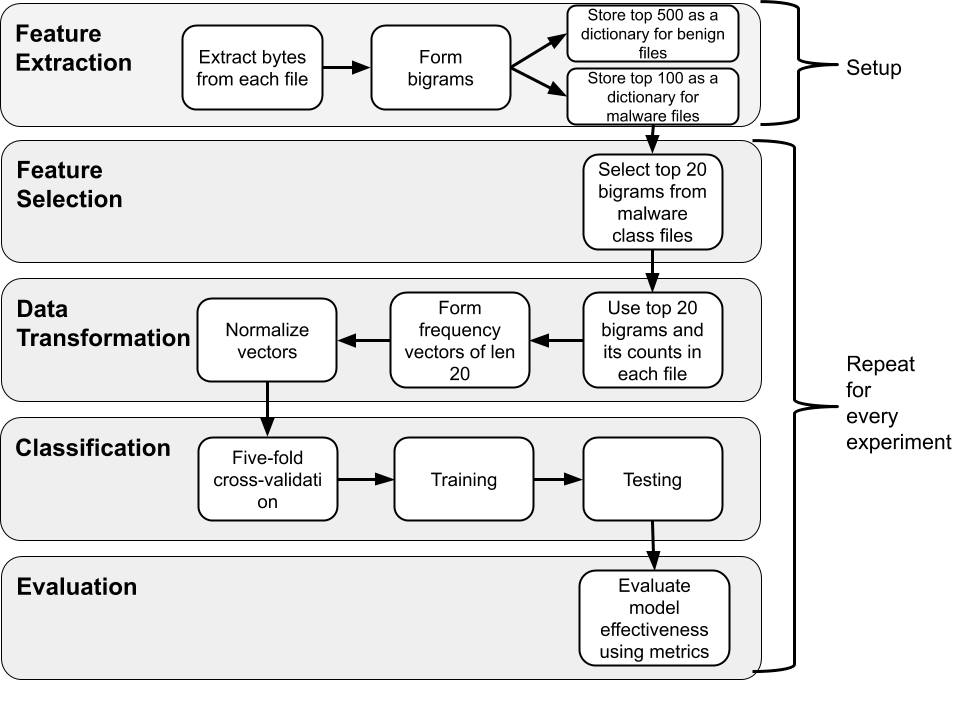}
\caption{Experiment Pipeline}\label{fig:experimentpipeline}
\end{figure}

In this section, we discuss the process of selecting features, training models, and scoring.
Figure~\ref{fig:experimentpipeline} illustrates the experiment 
pipeline. 

From the malware class, the top~20 bigrams are selected based on their frequencies. 
In cases where multiple families are combined, these multiple families belong to one 
class---the malware class---and the top bigrams are chosen from among these combined samples. 

From each sample in the benign and malware classes, a vector of frequencies corresponding 
to the top~20 bigrams is determined. These vectors are normalized to be row stochastic, which then 
form the feature vectors that are used for training and testing. Note that the set of feature
vectors changes with each combination of families, since we construct a new model in each case.

For the classification step, we use cross-validation to reduce potential bias 
and to maximize the number of experimental results.
We use a stratified $k$-fold technique where each fold maintains the 
class ratios from the dataset. 
Specifically, we use stratified five-fold cross-validation.

\section{Implementation}\label{chap:implementation}

In this section, we give a broad summary of the malware families and benign dataset
used in this research. Also, we briefly discuss the machine learning techniques that we have
employed in our experiments.

\subsection{Dataset}

As mentioned above, each family consists of~1000 samples. 
The benign set contains~1000 {\tt exe} files extracted from a fresh install 
on a Windows~10 laptop. Three of the malware families we consider, namely,
Winwebsec, Zeroaccess, and Zbot,
are from the Malicia dataset~\cite{nappa2015malicia}, while the remaining~17 malware families 
are taken from a big dataset collected using VirusShare as discussed in~\cite{bigdata}. 
This big dataset is almost half a 
terabyte and contains more than~500,000 malware samples in the form of 
labeled executable files. 
From this big dataset, we only considered families that each have at least~1000 samples. 

Table~\ref{tab:families} lists the~20 families used in this research, along with
the type of malware present in each family.  
Below, we briefly discuss each of these~20 malware families.

\begin{table}[!htb]
\begin{center}
\caption{Type of each malware family}\label{tab:families}
{\footnotesize
\begin{tabular}{cc|cc}\midrule\midrule
\textbf{Family} & \textbf{Type} & \textbf{Family}  & \textbf{Type} \\
\midrule
Adload~\cite{adload}      & Trojan Downloader            & Obfuscator~\cite{obfuscator}  & VirTool             \\
Agent~\cite{agent}       & Trojan        & OnLineGames~\cite{onlinegames} & Password Stealer            \\
Alureon~\cite{alureon}     & Trojan           & Rbot~\cite{rbot}        & Backdoor            \\
BHO~\cite{bho}         & Trojan          & Renos~\cite{renos}    & Trojan Downloader            \\
CeeInject~\cite{ceeinject}   & VirTool            & Startpage~\cite{startpage}   & Trojan            \\
Cycbot.G~\cite{cycbot}    & Backdoor            & Vobfus~\cite{vobfus}      & Worm            \\
DelfInject~\cite{delfinject}  & VirTool            & Vundo~\cite{vundo}       & Trojan Downloader            \\
FakeRean~\cite{fakerean}    & Rogue            & Winwebsec~\cite{winwebsec}   & Rogue            \\
Hotbar~\cite{hotbar}      & Adware            & Zbot~\cite{zbot}        & Password Stealer            \\
Lolyda.BF~\cite{lolyda}   & Password Stealer            & Zeroaccess~\cite{zeroaccess}  & Trojan Horse     \\
\midrule\midrule 
\end{tabular}
}
\end{center}
\end{table}

\begin{description}
\item[Adload] downloads an executable file, stores it remotely, executes the file, and disables 
proxy settings~\cite{adload}. 
\item[Agent] downloads trojans or other software from a remote server~\cite{agent}. 
\item[Alureon] exfiltrates usernames, passwords, credit card data, and 
other confidential data from an infected system~\cite{alureon}. 
\item[BHO] can perform a variety of actions, guided by an attacker~\cite{bho}. 
\item[CeeInject] uses advanced obfuscation to avoid being detected by antivirus software~\cite{ceeinject}. 
\item[Cycbot.G] connects to a remote server, exploits vulnerabilities, and spreads through backdoor 
ports~\cite{cycbot}. 
\item[DelfInject] sends usernames, passwords, and other personal 
and private information to an attacker~\cite{delfinject}. 
\item[FakeRean] pretends to scan the system, notifies the user of supposed issues, 
and asks the user to pay to clean the system~\cite{fakerean}. 
\item[Hotbar] is an adware that shows ads on webpages and installs additional adware~\cite{hotbar}. 
\item[Lolyda.BF] sends information from an infected system and monitors the system. It can share user 
credentials and network activity with an attacker~\cite{lolyda}. 
\item[Obfuscator] tries to obfuscate or hide itself to defeat malware detectors~\cite{obfuscator}.
\item[OnLineGames] steals login information of online games and tracks user 
keystroke activity~\cite{onlinegames}. 
\item[Rbot] gives control to attackers via a backdoor that can be used to access information or
launch attacks, and serves as a gateway to infect additional sites~\cite{rbot}.
\item[Renos] downloads software that claims the system has spyware and asks for a payment to 
remove the nonexistent spyware~\cite{renos}. 
\item[Startpage] changes the default browser homepage and may perform other
malicious activities~\cite{startpage}. 
\item[Vobfus] is a worm that downloads malware and spreads through USB drives or other 
removable drives~\cite{vobfus}. 
\item[Vundo] displays pop-up ads and may download files. It uses advanced techniques to 
defeat detection~\cite{vundo}.
\item[Winwebsec] displays alerts that ask the user for money to 
fix supposed issues~\cite{winwebsec}.
\item[Zbot] is installed through email and shares a user's personal information with attackers.
In addition, Zbot can disable a firewall~\cite{zbot}.
\item[Zeroaccess] is a trojan horse that downloads applications that click on ads,
thereby making money for the creator of the malware~\cite{zeroaccess}.
\end{description}

\subsection{Classification Techniques}

In this section, we briefly discuss the machine learning techniques used in our 
experiments. Specifically, we consider $k$-nearest neighbors, support vector machines, 
random forests, and multilayer perceptrons.

\subsubsection{$k$-Nearest Neighbors}

One of the simplest algorithms in machine learning is $k$-nearest neighbors (\kNN). 
In the scoring phase, \kNN\ consists of classifying based on the $k$ nearest samples
in the training set, typically using a simple majority vote. 
Since all computation is deferred to the scoring phase, \kNN\ 
is considered to be a ``lazy learner.''

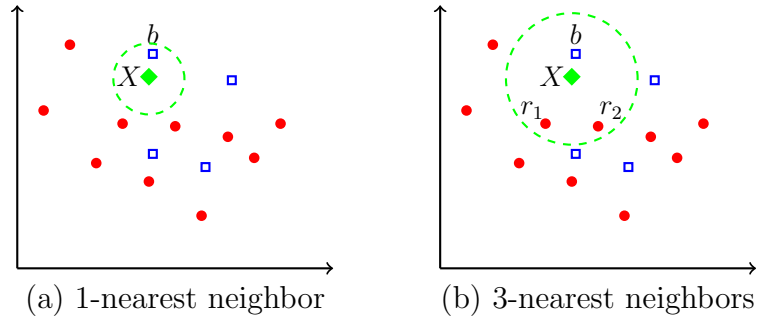
\begin{figure}[!htb]
\centering
  \begin{tabular}{ccc}
	    \begin{tikzpicture}[scale=0.7, every node/.style={scale=0.9}]
        
    \draw[thick,color=red,fill=red] (0.5,3) circle (0.08);
    \draw[thick,color=red,fill=red] (1.0,4.25) circle (0.08);
    \draw[thick,color=red,fill=red] (1.5,2.0) circle (0.08);
    \draw[thick,color=red,fill=red] (2.0,2.75) circle (0.08);%
    \draw[thick,color=red,fill=red] (2.5,1.65) circle (0.08);
    \draw[thick,color=red,fill=red] (3.0,2.7) circle (0.08);%
    \draw[thick,color=red,fill=red] (3.5,1.0) circle (0.08);
    \draw[thick,color=red,fill=red] (4.0,2.5) circle (0.08);
    \draw[thick,color=red,fill=red] (4.5,2.1) circle (0.08);
    \draw[thick,color=red,fill=red] (5.0,2.75) circle (0.08);

    \draw[thick,color=blue] (2.5,4.0) rectangle (2.65,4.15);%
    \node[color=black] at (2.575,4.45){$b$};
    \draw[thick,color=blue] (2.5,2.1) rectangle (2.65,2.25);
    \draw[thick,color=blue] (3.5,1.85) rectangle (3.65,2.0);
    \draw[thick,color=blue] (4.0,3.5) rectangle (4.15,3.65);

    \draw[thick,color=green,fill=green,rotate around={45:(2.5,3.5)}] (2.5,3.5) rectangle (2.7,3.7);
    \node[color=black] at (2.125,3.65){$X$};

    \draw[thick,color=green,dashed] (2.5,3.6) circle (0.675);

    \draw[thick,color=black,->] (0,0) -- (6,0); 
    \draw[thick,color=black,->] (0,0) -- (0,5); 
      

    \end{tikzpicture}
	& \ \ \ \ \ &
	    \begin{tikzpicture}[scale=0.7, every node/.style={scale=0.9}]
        
    \draw[thick,color=red,fill=red] (0.5,3) circle (0.08);
    \draw[thick,color=red,fill=red] (1.0,4.25) circle (0.08);
    \draw[thick,color=red,fill=red] (1.5,2.0) circle (0.08);
    \draw[thick,color=red,fill=red] (2.0,2.75) circle (0.08);%
    \node[color=black] at (1.75,3.0){$r_1$};
    \draw[thick,color=red,fill=red] (2.5,1.65) circle (0.08);
    \draw[thick,color=red,fill=red] (3.0,2.7) circle (0.08);%
    \node[color=black] at (3.25,3.0){$r_2$};
    \draw[thick,color=red,fill=red] (3.5,1.0) circle (0.08);
    \draw[thick,color=red,fill=red] (4.0,2.5) circle (0.08);
    \draw[thick,color=red,fill=red] (4.5,2.1) circle (0.08);
    \draw[thick,color=red,fill=red] (5.0,2.75) circle (0.08);

    \draw[thick,color=blue] (2.5,4.0) rectangle (2.65,4.15);%
    \node[color=black] at (2.575,4.45){$b$};
    \draw[thick,color=blue] (2.5,2.1) rectangle (2.65,2.25);
    \draw[thick,color=blue] (3.5,1.85) rectangle (3.65,2.0);
    \draw[thick,color=blue] (4.0,3.5) rectangle (4.15,3.65);

    \draw[thick,color=green,fill=green,rotate around={45:(2.5,3.5)}] (2.5,3.5) rectangle (2.7,3.7);
    \node[color=black] at (2.125,3.65){$X$};

    \draw[thick,color=green,dashed] (2.5,3.6) circle (1.25);

    \draw[thick,color=black,->] (0,0) -- (6,0); 
    \draw[thick,color=black,->] (0,0) -- (0,5); 
      

    \end{tikzpicture}
	\\
	(a) 1-nearest neighbor & & (b) 3-nearest neighbors
  \end{tabular}	
\caption{Examples of \kNN\ classification~\cite{stamp2017introduction}}\label{fig:knn}
\end{figure}

Figure~\ref{fig:knn} shows examples of \kNN, where the training 
data consists of two classes, represented by the open blue squares and the
solid red circles, with the green diamond (the point labeled~$X$) being
a point that we want to classify. 
Figure~\ref{fig:knn}~(a), shows that if we use the 1-nearest
neighbor, we would classify the green diamond 
as being of same type as the open blue squares, whereas Figure~\ref{fig:knn}~(b)
shows that~$X$ would be classified as the solid red circle type if using the
3-nearest neighbors. 

\subsubsection{Support Vector Machines}

Support vector machines (SVM) are a class of 
supervised learning methods that are based on four major ideas,
namely, a separating hyperplane, maximizing the ``margin'' (i.e., separation between classes), 
working in a higher-dimensional space, and the so-called kernel trick. The goal in SVM is to
use a hyperplane to separate labeled data into two classes. If it exists, such a hyperplane 
is chosen to maximize the margin~\cite{stamp2017introduction}.
  
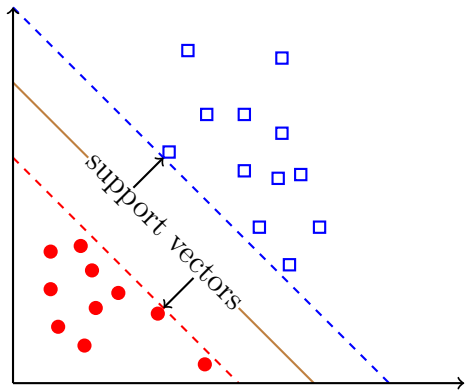
\begin{figure}[!htb]
\centering
    \begin{tikzpicture}[scale=1.0]
    
    \draw[thick,color=blue] (4,2) rectangle (4.15,2.15);
    \draw[thick,color=blue] (3.5,4.25) rectangle (3.65,4.4);
    \draw[thick,color=blue] (3.2,2.0) rectangle (3.35,2.15);
    \draw[thick,color=blue] (3.0,2.75) rectangle (3.15,2.9);
    \draw[thick,color=blue] (3.45,2.65) rectangle (3.6,2.8);
    \draw[thick,color=blue] (3.75,2.7) rectangle (3.9,2.85);
    \draw[thick,color=blue] (3.5,3.25) rectangle (3.65,3.4);
    \draw[thick,color=blue] (3.0,3.5) rectangle (3.15,3.65);
    \draw[thick,color=blue] (2,3) rectangle (2.15,3.15);
    \draw[thick,color=blue] (2.5,3.5) rectangle (2.65,3.65);
    \draw[thick,color=blue] (2.25,4.35) rectangle (2.4,4.5);
    \draw[thick,color=blue] (3.6,1.5) rectangle (3.75,1.65);
    
    \draw[thick,color=red,fill=red] (1.4,1.2) circle (0.08);
    \draw[thick,color=red,fill=red] (0.6,0.75) circle (0.08);
    \draw[thick,color=red,fill=red] (0.95,0.5) circle (0.08);
    \draw[thick,color=red,fill=red] (1.925,0.925) circle (0.08);
    \draw[thick,color=red,fill=red] (0.5,1.25) circle (0.08);
    \draw[thick,color=red,fill=red] (1.1,1.0) circle (0.08);
    \draw[thick,color=red,fill=red] (2.55,0.25) circle (0.08);
    \draw[thick,color=red,fill=red] (1.05,1.5) circle (0.08);
    \draw[thick,color=red,fill=red] (0.9,1.825) circle (0.08);
    \draw[thick,color=red,fill=red] (0.5,1.75) circle (0.08);
    

    
    \draw[thick,color=brown] (0,4) -- (1,3); 
    \draw[thick,color=brown] (3,1) -- (4,0); 
    \draw[thick,dashed,color=blue] (0,5) -- (5,0); 
    \draw[thick,dashed,color=red] (0,3) -- (3,0); 

    \draw[thick,color=black,->] (2.4,1.4) -- (2,1); 
    \draw[thick,color=black,->] (1.6,2.6) -- (2,3); 
    

    \node[rotate=-45] at (2,2){support vectors};
    
     \draw[thick,color=black,->] (0,0) -- (6,0); 
     \draw[thick,color=black,->] (0,0) -- (0,5); 
   
    \end{tikzpicture}
\caption{Support vectors in SVM~\cite{stamp2017introduction}}\label{fig:svm}
\end{figure}

An example of a trained SVM is illustrated in Figure~\ref{fig:svm}. Note that the
points that actually minimize the distance to the separating hyperplane
correspond to support vectors. In general, the number of support vectors
will be small relative to the number of training data points,
and this is the key to the efficiency of SVM in the classification phase.

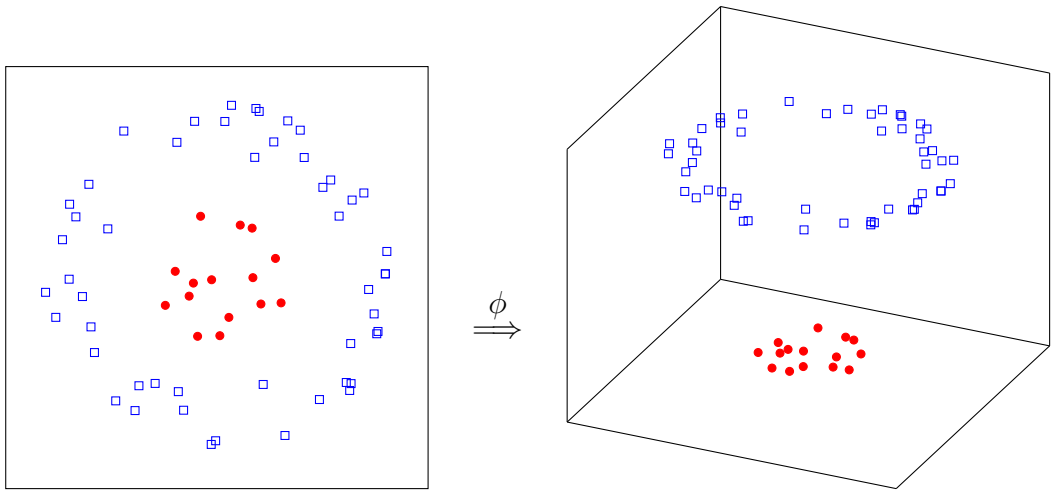
\begin{figure}[!htb]
\centering
  \begin{tabular}{ccc}
  \begin{tikzpicture} 
    \begin{axis}[width=0.45\textwidth,height=0.45\textwidth,xmin=-3,xmax=3,ymin=-3,ymax=3] 
      \pgfplotsset{ticks=none}
      
       \addplot[color=red, mark=*,only marks,mark size=1.5] coordinates { 
(0.331578, 0.746603)
(-0.394443, -0.263352)
(0.625403, -0.375284)
(0.911628, -0.359273)
(0.500405, 0.703271)
(-0.231736, 0.872177)
(0.170166, -0.565556)
(-0.274232, -0.834498)
(-0.334136, -0.078180)
(-0.592072, 0.089388)
(-0.075179, -0.030616)
(0.041852, -0.825840)
(-0.732711, -0.394369)
(0.511456, -0.000657)
(0.832060, 0.272873)
      };
      \addplot[color=blue, mark=square,fill=blue,only marks,mark size=1.5] coordinates { 
(2.390971, 0.052211)
(-1.437509, -1.752351)
(1.886234, -1.604676)
(2.395163, 0.054876)
(2.288330, -0.765636)
(0.807904, 1.929355)
(0.207518, 2.449499)
(1.900863, -0.937399)
(-0.548196, -1.621026)
(-1.322488, 2.083358)
(1.006897, 2.228109)
(1.735974, 0.874161)
(1.838050, -1.491165)
(0.600720, 2.362412)
(1.456580, -1.733536)
(-0.316657, 2.221768)
(2.087640, 1.203409)
(1.616520, 1.387743)
(-2.433935, -0.209267)
(-1.788833, -0.700340)
(-0.877470, -1.504627)
(-1.107672, -1.537246)
(1.184006, 2.095965)
(-0.080668, -2.371692)
(2.414325, 0.371814)
(0.554228, 2.404566)
(1.505664, 1.283217)
(-1.162894, -1.890363)
(-2.098915, -0.022764)
(1.920753, 1.102422)
(-2.288927, -0.565147)
(1.905982, -1.506967)
(2.233564, -0.516066)
(-2.003422, 0.863837)
(-1.910277, -0.269332)
(0.966902, -2.244612)
(-0.568292, 1.923142)
(0.112558, 2.219818)
(-1.549785, 0.692908)
(-2.193879, 0.538788)
(-1.818708, 1.326454)
(2.270850, -0.798137)
(-2.092595, 1.043138)
(2.155481, -0.168913)
(-0.017986, -2.319378)
(1.239888, 1.707677)
(-0.473226, -1.885332)
(0.537416, 1.709329)
(-1.739303, -1.065284)
(0.657574, -1.519209)
      };
    \end{axis}
  \end{tikzpicture}  
  & \raisebox{0.85in}{${{\displaystyle\phi} \atop {\displaystyle\implies}}$} &
    \begin{tikzpicture}
    \begin{axis}[width=0.5\textwidth,height=0.5\textwidth,xmin=-3,xmax=3,ymin=-3,ymax=3,zmin=0,zmax=5] 
      \pgfplotsset{ticks=none}
      \addplot3[color=red, mark=*,only marks,mark size=1.5] coordinates { 
(0.331578, 0.746603,0.6)
(-0.394443, -0.263352,0.6)
(0.625403, -0.375284,0.6)
(0.911628, -0.359273,0.6)
(0.500405, 0.703271,0.6)
(-0.231736, 0.872177,0.6)
(0.170166, -0.565556,0.6)
(-0.274232, -0.834498,0.6)
(-0.334136, -0.078180,0.6)
(-0.592072, 0.089388,0.6)
(-0.075179, -0.030616,0.6)
(0.041852, -0.825840,0.6)
(-0.732711, -0.394369,0.6)
(0.511456, -0.000657,0.6)
(0.832060, 0.272873,0.6)
      };
      \addplot3[color=blue, mark=square,fill=blue,only marks,mark size=1.5] coordinates { 
(2.390971, 0.052211,4.0)
(-1.437509, -1.752351,4.0)
(1.886234, -1.604676,4.0)
(2.395163, 0.054876,4.0)
(2.288330, -0.765636,4.0)
(0.807904, 1.929355,4.0)
(0.207518, 2.449499,4.0)
(1.900863, -0.937399,4.0)
(-0.548196, -1.621026,4.0)
(-1.322488, 2.083358,4.0)
(1.006897, 2.228109,4.0)
(1.735974, 0.874161,4.0)
(1.838050, -1.491165,4.0)
(0.600720, 2.362412,4.0)
(1.456580, -1.733536,4.0)
(-0.316657, 2.221768,4.0)
(2.087640, 1.203409,4.0)
(1.616520, 1.387743,4.0)
(-2.433935, -0.209267,4.0)
(-1.788833, -0.700340,4.0)
(-0.877470, -1.504627,4.0)
(-1.107672, -1.537246,4.0)
(1.184006, 2.095965,4.0)
(-0.080668, -2.371692,4.0)
(2.414325, 0.371814,4.0)
(0.554228, 2.404566,4.0)
(1.505664, 1.283217,4.0)
(-1.162894, -1.890363,4.0)
(-2.098915, -0.022764,4.0)
(1.920753, 1.102422,4.0)
(-2.288927, -0.565147,4.0)
(1.905982, -1.506967,4.0)
(2.233564, -0.516066,4.0)
(-2.003422, 0.863837,4.0)
(-1.910277, -0.269332,4.0)
(0.966902, -2.244612,4.0)
(-0.568292, 1.923142,4.0)
(0.112558, 2.219818,4.0)
(-1.549785, 0.692908,4.0)
(-2.193879, 0.538788,4.0)
(-1.818708, 1.326454,4.0)
(2.270850, -0.798137,4.0)
(-2.092595, 1.043138,4.0)
(2.155481, -0.168913,4.0)
(-0.017986, -2.319378,4.0)
(1.239888, 1.707677,4.0)
(-0.473226, -1.885332,4.0)
(0.537416, 1.709329,4.0)
(-1.739303, -1.065284,4.0)
(0.657574, -1.519209,4.0)
      };
    \end{axis}
  \end{tikzpicture}  
  \end{tabular}
\caption{The function~$\phi$ illustrates the kernel trick~\cite{stamp2017introduction}}\label{fig:svm2}
\end{figure}

Of course, there is no assurance that the training data will be linearly separable. 
In such cases,
a nonlinear kernel function can be embedded into the SVM process
in such a way that the input data is, in effect, 
transformed to a higher dimensional ``feature space.'' 
In this higher dimensional space, it is far more likely that the 
transformed data will be linearly separable. This is the essence of the 
kernel trick---an example of which
is illustrated in Figure~\ref{fig:svm2}. That we can transform our training data 
in such a manner is not surprising, 
but the fact that we can do so 
without paying any significant penalty in terms of computational efficiency 
makes the kernel trick a very powerful ``trick'' indeed. However,
the kernel function must be specified by the user, and selecting a (near) optimal
kernel can be challenging.

\subsubsection{Random Forest}

A random forest (RF) generalizes a simple decision tree algorithm. 
A decision tree is constructed by building a tree, based on 
features from the training data. It is easy to construct such trees,
and trivial to classify samples once a tree has been constructed. However, 
decision trees tend to overfit the input data. 

An RF combines multiple decision trees to generalize the training data. 
To do so, RFs use different subsets of the training data as well as 
different subsets of features, a process known as bagging~\cite{stamp2017introduction}.
A simple majority vote of the decision trees comprising the RF is
typically used for classification~\cite{liaw2002classification}.

\subsubsection{Multilayer Perceptron}

\begin{figure}[!htb]
\centering
    \begin{tikzpicture}[scale=0.9]
    
    \draw[thick,color=green] (3.0,8.5) circle (0.575);
    \draw[thick,color=green] (5.5,8.5) circle (0.575);

    \draw[thick,color=blue] (1.25,5.5) rectangle (2.25,6.5);
    \draw[thick,color=blue] (3.75,5.5) rectangle (4.75,6.5);
    \draw[thick,color=blue] (6.25,5.5) rectangle (7.25,6.5);
    
    \draw[thick,color=blue] (0.0,2.5) rectangle (1.0,3.5);
    \draw[thick,color=blue] (2.5,2.5) rectangle (3.5,3.5);
    \draw[thick,color=blue] (5.0,2.5) rectangle (6.0,3.5);
    \draw[thick,color=blue] (7.5,2.5) rectangle (8.5,3.5);
    
    \draw[thick,color=red,rotate around={45:(1.75,0.5)}] (1.25,0.0) rectangle (2.25,1.0);
    \draw[thick,color=red,rotate around={45:(4.25,0.5)}] (3.75,0.0) rectangle (4.75,1.0);
    \draw[thick,color=red,rotate around={45:(6.75,0.5)}] (6.25,0.0) rectangle (7.25,1.0);

    \draw[thick,color=green,->] (3,7.9) -- (1.75,6.5);
    \draw[thick,color=green,->] (3,7.9) -- (4.25,6.5);
    \draw[thick,color=green,->] (3,7.9) -- (6.75,6.5);
    \draw[thick,color=green,->] (5.5,7.9) -- (1.75,6.5);
    \draw[thick,color=green,->] (5.5,7.9) -- (4.25,6.5);
    \draw[thick,color=green,->] (5.5,7.9) -- (6.75,6.5);

    \draw[thick,color=blue,->] (0.5,2.5) -- (1.4,0.85);
    \draw[thick,color=blue,->] (0.5,2.5) -- (3.9,0.85);
    \draw[thick,color=blue,->] (0.5,2.5) -- (6.4,0.85);
    \draw[thick,color=blue,->] (3.0,2.5) -- (2.1,0.85);
    \draw[thick,color=blue,->] (3.0,2.5) -- (3.9,0.85);
    \draw[thick,color=blue,->] (3.0,2.5) -- (6.4,0.85);
    \draw[thick,color=blue,->] (5.5,2.5) -- (2.1,0.85);
    \draw[thick,color=blue,->] (5.5,2.5) -- (4.6,0.85);
    \draw[thick,color=blue,->] (5.5,2.5) -- (6.4,0.85);
    \draw[thick,color=blue,->] (8.0,2.5) -- (2.1,0.85);
    \draw[thick,color=blue,->] (8.0,2.5) -- (4.6,0.85);
    \draw[thick,color=blue,->] (8.0,2.5) -- (7.1,0.85);

    \draw[thick,color=blue,->] (1.75,5.5) -- (0.5,3.5);
    \draw[thick,color=blue,->] (1.75,5.5) -- (3.0,3.5);
    \draw[thick,color=blue,->] (1.75,5.5) -- (5.5,3.5);
    \draw[thick,color=blue,->] (1.75,5.5) -- (8.0,3.5);

    \draw[thick,color=blue,->] (4.25,5.5) -- (0.5,3.5);
    \draw[thick,color=blue,->] (4.25,5.5) -- (3.0,3.5);
    \draw[thick,color=blue,->] (4.25,5.5) -- (5.5,3.5);
    \draw[thick,color=blue,->] (4.25,5.5) -- (8.0,3.5);

    \draw[thick,color=blue,->] (6.75,5.5) -- (0.5,3.5);
    \draw[thick,color=blue,->] (6.75,5.5) -- (3.0,3.5);
    \draw[thick,color=blue,->] (6.75,5.5) -- (5.5,3.5);
    \draw[thick,color=blue,->] (6.75,5.5) -- (8.0,3.5);

    \draw[thick,color=red,->] (1.75,-0.2) -- (1.75,-1.2);
    \draw[thick,color=red,->] (4.25,-0.2) -- (4.25,-1.2);
    \draw[thick,color=red,->] (6.75,-0.2) -- (6.75,-1.2);

    \node at (3.0,8.5) {$X_1$};
    \node at (5.5,8.5) {$X_2$};

    \node at (1.75,6.0) {$f_1$};
    \node at (4.25,6.0) {$f_1$};
    \node at (6.75,6.0) {$f_1$};

    \node at (0.5,3.0) {$f_2$};
    \node at (3.0,3.0) {$f_2$};
    \node at (5.5,3.0) {$f_2$};
    \node at (8.0,3.0) {$f_2$};

    \node at (1.75,0.5) {$g$};
    \node at (4.25,0.5) {$g$};
    \node at (6.75,0.5) {$g$};

    \node at (1.75,-1.65) {$Y_1$};
    \node at (4.25,-1.65) {$Y_2$};
    \node at (6.75,-1.65) {$Y_3$};
    
    \node at (10.5,8.5) {Input layer};
    \node at (10.5,6.0) {1st hidden layer};
    \node at (10.5,3.0) {2nd hidden layer};
    \node at (10.5,0.5) {Output layer};
    \node at (10.5,-1.65) {Output};

    \end{tikzpicture}
\caption{MLP with two hidden layers~\cite{stamp2017introduction}}\label{fig:mini_MLP}
\end{figure}
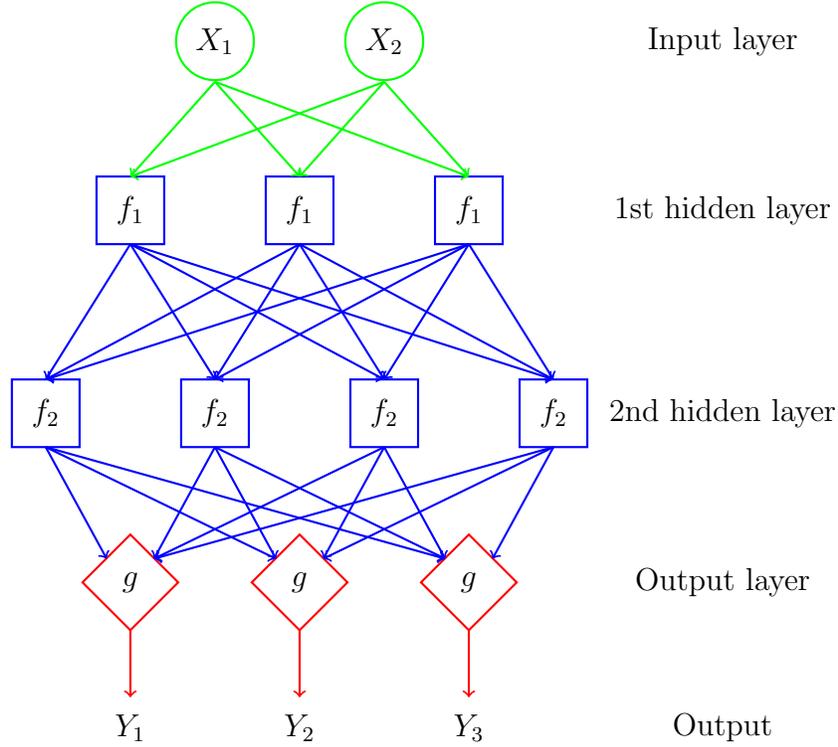

Neural networks can be viewed as modeling neurons in the brain. 
A perceptron is a feedforward type of artificial neuron that has an input layer and an output layer.
While conceptually simple, a perceptron is limited to a linear decision boundary, much
like a linear SVM. The equivalent of the kernel trick for perceptrons is the multilayer perceptron (MLP),
which includes one or more hidden layers between the input and the output. An example
of an MLP with two hidden layers is given in Figure~\ref{fig:mini_MLP}. The number 
of layers, the number of neurons (i.e., functions) in each layer, and the functions themselves
must be specified as part of an MLP architecture.
Each edge in an MLP graph represents a weight that is determined 
via training. Backpropagation, which is a gradient descent technique,
provides an efficient means of training an MLP~\cite{stamp2017introduction}.

Due to the hidden layers, an MLP is not restricted to a linear decision boundary, 
which is very much analogous to an SVM with a nonlinear kernel function. The advantage of
an MLP over a nonlinear SVM is that we do not specify an explicit kernel function.
With an MLP, it is as if an optimal kernel function---in the SVM sense---is determined during training.
However, more training data and computational effort is needed to train 
an MLP (as compared to a nonlinear SVM) since more parameters must be determined.
Other potential issues with MLPs include that they are not robust against feature scaling 
and that there are multiple local minima that can be found during training~\cite{scikitlearn}.

\subsubsection{Discussion}

There are some interesting connections between pairs of the ML techniques that we have chosen 
for our experiments. As mentioned above, an MLP can be viewed as being analogous to 
an SVM, where the equivalent of the kernel function is learned as part of the training.

Perhaps somewhat less obvious is the connection between \kNN\ and RF.
Both of these techniques are, in fact, neighborhood-based algorithms. 
To see that such is the case, suppose that we are given a labeled training 
set~$(X_i,z_i)$, for~$i=1,2,\ldots,n$, with each~$z_i\in\{-1,+1\}$. Further,
suppose that we are given a sample~$X$ to classify.

For~\kNN, we can define the weight function
$$
  W_{k}(X_i,X) = \left\{
  \begin{array}{ll}
  1 & \mbox{if $X_i$ is one of the $k$ nearest neighbors to $X$}\\
  0 & \mbox{otherwise} .
  \end{array}
  \right.
$$
We could then compute 
$$
  \mbox{score}_k(X) = \sum_{i=1}^n z_i\, W_{k}(X_i,X)
$$
and we would classify~$X$ as type~$+1$ provided that~$\mbox{score}_k(X) > 0$,
and as type~$-1$ if~$\mbox{score}_k(X) < 0$.

For this same training set and sample~$X$, suppose that we use
a decision tree instead. Then we can define
$$
  W_{t}(X_i,X) = \left\{
  \begin{array}{ll}
  1 & \mbox{if $X_i$ is on the same leaf node as $X$}\\
  0 & \mbox{otherwise} .
  \end{array}
  \right.
$$
With this weight function, we compute 
\begin{equation}\label{eq:mini_score_RF}
  \mbox{score}_{t}(X) = \sum_{i=1}^n z_i\, W_{t}(X_i,X)
\end{equation}
and the decision tree classifies~$X$ as type~$+1$,
provided that~$\mbox{score}_t(X) > 0$, and of type~$-1$ 
if~$\mbox{score}_t(X) < 0$.

From this perspective, we see 
that~\kNN\ and decision trees are both neighborhood-based 
classification algorithms, but with different neighborhood structures. Since an RF 
is a collection of decision trees, a random forest is also a neighborhood-based
classification technique, but with a relatively complex
neighborhood structure~\cite{stamp2017introduction}.

Based on this discussion, we would generically expect SVM and MLP to perform
in qualitatively similar ways. And, we would also expect \kNN\ and RF to
perform somewhat similarly. Thus, by choosing these four specific algorithms,
we obtain something of a ``higher order'' check on our results. That is, we 
expect to see more similarity between the performance of SVM and MLP, 
as well as RF and \kNN, as compared to the other pairs, such as SVM and RF.

\subsection{Evaluation Metrics}

Accuracy is defined as the number of correct classifications divided by the 
total number of samples and is calculated as 
$$
  \textrm{accuracy} = \frac{\mbox{TP}+\mbox{TN}}{\mbox{P}+\mbox{N}}
$$
where~TP (true positive) is the number of samples correctly classified as positive 
and~TN (true negative) is the number of samples correctly classified as negative. 
Also, we use~P for the total number of positive samples, and~N for the total number of negative 
samples. 

In our experiments, as we add more families, the datasets become progressively more imbalanced.
To negate the effect of this imbalance, we use balanced accuracy to evaluate 
our experimental results~\cite{stamp2017introduction}. 
The balanced accuracy is computed by simply weighting the positive and negative sets
equally, that is
$$
  \textrm{balanced accuracy} = \frac{1}{2}\biggl(\frac{\mbox{TP}}{\mbox{P}} 
  	+ \frac{\mbox{TN}}{\mbox{N}}\biggr) .
$$
In the remainder of this paper, when we mention accuracy, 
we are referring to the balanced accuracy. 

\subsection{Implementation Details}

For each malware sample,
a dictionary containing the top~100 bigrams is constructed,
while for benign samples, 
the dictionary contains the top~500 bigrams. In addition, for
each of the~20 malware families under consideration, we construct
a dictionary consisting of the top~1500 bigrams from the overall family. 
These dictionaries make it possible to efficiently determine
our bigram features for any combination of families.

At level~$N$, we consider
$$
  \min\left(100, {20\choose N}\right) 
$$
models. In cases where~$100<{20\choose N}$,
the families used to construct each model
are chosen randomly (with replacement) from our set of~20
families. 

The \texttt{shuffle} function from the \texttt{random} module in Python~\cite{random} is used to randomly 
select a subset of families. This function uses the Fisher-Yates shuffle, which is an 
unbiased algorithm.  The algorithm runs through the list of combinations in reverse order and 
randomly picks an element to exchange. After the shuffle, the first~100 combinations are chosen. 

The popular \texttt{scikit-learn} library~\cite{scikitlearn} is used to generate
our SVM, \kNN, random forest, and 
MLP models. The \texttt{model\_selection} module is used to perform~5-fold cross-validation, 
and the \texttt{metrics} module calculates the balanced accuracy. A linear SVM model
is used in our SVM experiments. The neighbor size for \kNN\ is set to~$k=5$ and 
for the random forest, we use~10 as the number of estimators and~10 for the maximum depth. 
The \texttt{lbfgs} solver is used as a parameter for MLP. We experimented with
other combinations of parameters and found these to perform best.
We briefly revisit the issue of parameter selection in the next section.

\section{Results and Analysis}\label{chap:results}

In this section, we present our experimental results and provide some
analysis of these results. 
We begin by discussing our binary classification results for
individual family models, that is, the cases where we classify samples from 
a specific family, together with the benign class, as either malware or benign. 
Then we discuss the results of multifamily models, 
where we combine multiple malware families into the malware class and 
perform analogous binary classification experiments.

\subsection{Individual Family Models}

In this section, we list accuracies obtained for individual families for each of the
machine learning algorithms under consideration.
That is, we analyze the overall effectiveness of classifying one specific family 
as benign or malware. These are our level~1 experiments.

\subsubsection{Support Vector Machines}

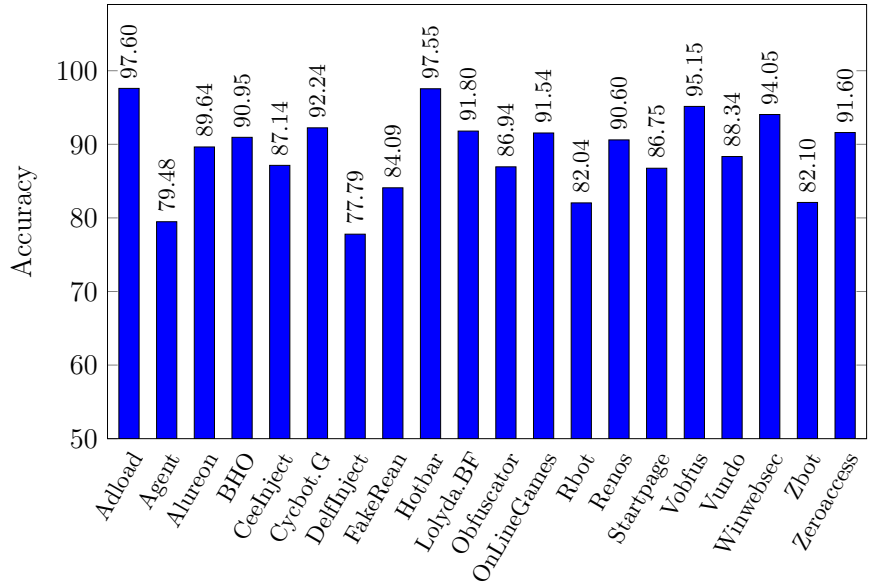
\begin{figure}[!htb]
\centering
\begin{tikzpicture}[scale=0.9, every node/.style={scale=1.0}]
    \begin{axis}[
        width  = 0.8*\textwidth,
        height = 8cm,
        ymin=50.0,ymax=109.0,
        ytick={50,60,70,80,90,100},
        major x tick style = transparent,
        ybar=4*\pgflinewidth,
        bar width=8.5pt,
        ylabel = {Accuracy},
        symbolic x coords={Adload,
				      Agent,
				      Alureon,
				      BHO,
				      CeeInject,
				      Cycbot.G,
				      DelfInject,
				      FakeRean,
				      Hotbar,
				      Lolyda.BF,
				      Obfuscator,
				      OnLineGames,
				      Rbot,
				      Renos,
				      Startpage,
				      Vobfus,
				      Vundo,
				      Winwebsec,
				      Zbot,
				      Zeroaccess},
	y tick label style={
    		/pgf/number format/.cd,
   		fixed,
   		fixed zerofill,
    		precision=0},
        xtick = data,
        x tick label style={
        		rotate=60,
		font=\footnotesize,
		anchor=north east,
		inner sep=0mm},
        nodes near coords,
        every node near coord/.append style={rotate=90, anchor=west, font=\footnotesize},
        enlarge x limits=0.03,
        legend cell align=left,
        legend style={
                at={(0.05,0)},
                anchor=south,
                column sep=1ex
        }
    ]
\addplot[fill=blue,opacity=1.00] 
coordinates {
(Adload,97.60)
(Agent,79.48)
(Alureon,89.64)
(BHO,90.95)
(CeeInject,87.14)
(Cycbot.G,92.24)
(DelfInject,77.79)
(FakeRean,84.09)
(Hotbar,97.55)
(Lolyda.BF,91.80)
(Obfuscator,86.94)
(OnLineGames,91.54)
(Rbot,82.04)
(Renos,90.60)
(Startpage,86.75)
(Vobfus,95.15)
(Vundo,88.34)
(Winwebsec,94.05)
(Zbot,82.10)
(Zeroaccess,91.60)
};
\end{axis}
\end{tikzpicture}
\caption{SVM individual family results}\label{fig:ind_SVM} 
\end{figure}

Figure~\ref{fig:ind_SVM} shows the results of classifying individual families using SVM. 
The average balanced accuracy at level~1 for this algorithm is~88.88\%. 
We see that two families (Agent and DelfInject) are far below this average, 
while three families (Vobfus, Adload, and Hotbar) are far above the average. 
For these experiments, we observe that an SVM performs poorly on some of the
families, with the range between the highest and the lowest accuracy being~19.81\%.
This is consistent with the SVM results for several of these same families in~\cite{Mayuri}.

\subsubsection{$k$-Nearest Neighbors}

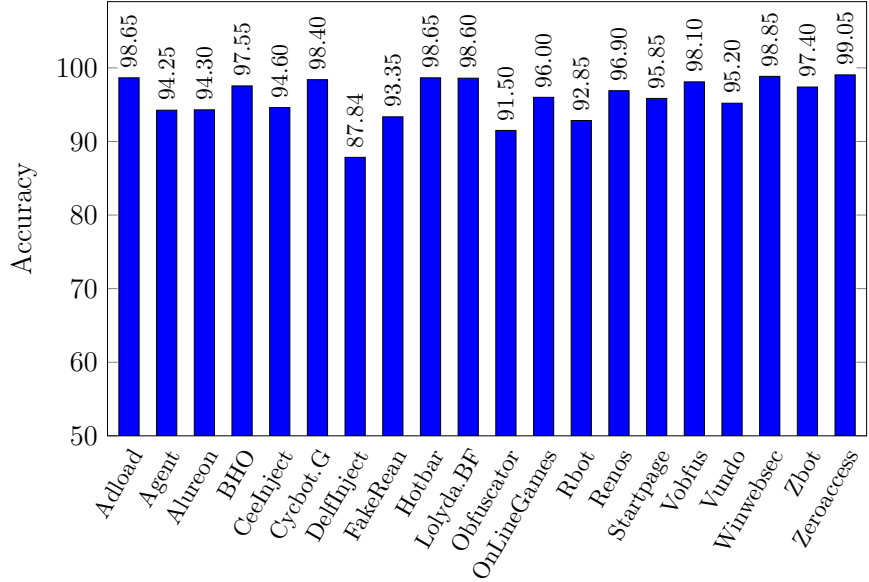
\begin{figure}[!htb]
\centering
\begin{tikzpicture}[scale=0.9, every node/.style={scale=1.0}]
    \begin{axis}[
        width  = 0.8*\textwidth,
        height = 8cm,
        ymin=50.0,ymax=109.0,
        ytick={50,60,70,80,90,100},
        major x tick style = transparent,
        ybar=4*\pgflinewidth,
        bar width=8.5pt,
        ylabel = {Accuracy},
        symbolic x coords={Adload,
				      Agent,
				      Alureon,
				      BHO,
				      CeeInject,
				      Cycbot.G,
				      DelfInject,
				      FakeRean,
				      Hotbar,
				      Lolyda.BF,
				      Obfuscator,
				      OnLineGames,
				      Rbot,
				      Renos,
				      Startpage,
				      Vobfus,
				      Vundo,
				      Winwebsec,
				      Zbot,
				      Zeroaccess},
	y tick label style={
    		/pgf/number format/.cd,
   		fixed,
   		fixed zerofill,
    		precision=0},
        xtick = data,
        x tick label style={
        		rotate=60,
		font=\footnotesize,
		anchor=north east,
		inner sep=0mm},
        nodes near coords,
        every node near coord/.append style={rotate=90, anchor=west, font=\footnotesize},
        enlarge x limits=0.03,
        legend cell align=left,
        legend style={
                at={(0.05,0)},
                anchor=south,
                column sep=1ex
        }
    ]
\addplot[fill=blue,opacity=1.00] 
coordinates {
(Adload,98.65)
(Agent,94.25)
(Alureon,94.30)
(BHO,97.55)
(CeeInject,94.60)
(Cycbot.G,98.40)
(DelfInject,87.84)
(FakeRean,93.35)
(Hotbar,98.65)
(Lolyda.BF,98.60)
(Obfuscator,91.50)
(OnLineGames,96.00)
(Rbot,92.85)
(Renos,96.90)
(Startpage,95.85)
(Vobfus,98.10)
(Vundo,95.20)
(Winwebsec,98.85)
(Zbot,97.40)
(Zeroaccess,99.05)
};
\end{axis}
\end{tikzpicture}
\caption{\kNN\ individual family results}\label{fig:ind_kNN} 
\end{figure}

Figure~\ref{fig:ind_kNN} gives the results for classifying individual families using \kNN. 
The average balanced accuracy at level~1 using \kNN\ is~95.87\%,
which is much higher than for SVM.
We note that nine families are below the average, with DelfInject far below,
while six families (Vobfus, Lolyda.BF, Adload, Hotbar, Zeroaccess, and Winwebsec) 
have accuracies greater than~98\%. 
The range in accuracy is~11.21\%, which is narrower than the SVM results. 
Overall, \kNN\ easily outperforms our SVM experiments.  

\subsubsection{Random Forest}

\begin{figure}[!htb]
\centering
\begin{tikzpicture}[scale=0.9, every node/.style={scale=1.0}]
    \begin{axis}[
        width  = 0.8*\textwidth,
        height = 8cm,
        ymin=50.0,ymax=109.0,
        ytick={50,60,70,80,90,100},
        major x tick style = transparent,
        ybar=4*\pgflinewidth,
        bar width=8.5pt,
        ylabel = {Accuracy},
        symbolic x coords={Adload,
				      Agent,
				      Alureon,
				      BHO,
				      CeeInject,
				      Cycbot.G,
				      DelfInject,
				      FakeRean,
				      Hotbar,
				      Lolyda.BF,
				      Obfuscator,
				      OnLineGames,
				      Rbot,
				      Renos,
				      Startpage,
				      Vobfus,
				      Vundo,
				      Winwebsec,
				      Zbot,
				      Zeroaccess},
	y tick label style={
    		/pgf/number format/.cd,
   		fixed,
   		fixed zerofill,
    		precision=0},
        xtick = data,
        x tick label style={
        		rotate=60,
		font=\footnotesize,
		anchor=north east,
		inner sep=0mm},
        nodes near coords,
        every node near coord/.append style={rotate=90, anchor=west, font=\footnotesize},
        enlarge x limits=0.03,
        legend cell align=left,
        legend style={
                at={(0.05,0)},
                anchor=south,
                column sep=1ex
        }
    ]
\addplot[fill=blue,opacity=1.00] 
coordinates {
(Adload,99.35)
(Agent,95.90)
(Alureon,98.00)
(BHO,99.20)
(CeeInject,97.45)
(Cycbot.G,99.35)
(DelfInject,95.60)
(FakeRean,96.65)
(Hotbar,99.35)
(Lolyda.BF,99.65)
(Obfuscator,96.10)
(OnLineGames,98.15)
(Rbot,96.90)
(Renos,98.15)
(Startpage,98.75)
(Vobfus,99.45)
(Vundo,97.95)
(Winwebsec,99.95)
(Zbot,99.25)
(Zeroaccess,99.45)
};
\end{axis}
\end{tikzpicture}
\caption{Random forest individual family results}\label{fig:ind_RF} 
\end{figure}
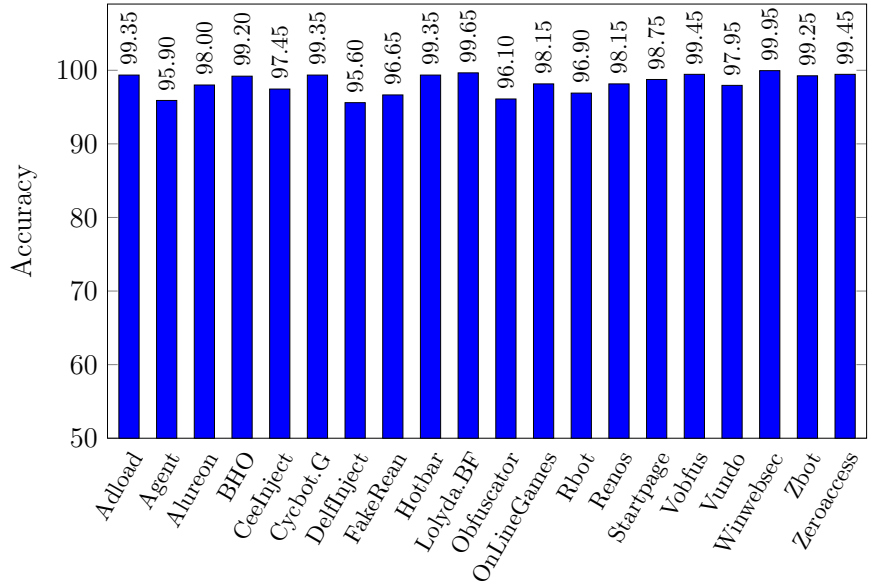

Figure~\ref{fig:ind_RF} contains results for classifying individual families using the 
random forest algorithm. 
In this case, the average balanced accuracy at level~1 using RF is~98.23\%, 
which is even better than \kNN. We see that~10 families are below the average 
accuracy. DelfInject and Agent have the lowest accuracies, but are less than 3\%\ below the average. 
Winwebsec, the family with the highest accuracy, is at~99.95\%. The range in accuracy is 
a narrow~4.35\%, indicating that RF consistently yields strong results. 

\subsubsection{Multilayer Perceptron}

\begin{figure}[!htb]
\centering
\begin{tikzpicture}[scale=0.9, every node/.style={scale=1.0}]
    \begin{axis}[
        width  = 0.8*\textwidth,
        height = 8cm,
        ymin=50.0,ymax=109.0,
        ytick={50,60,70,80,90,100},
        major x tick style = transparent,
        ybar=4*\pgflinewidth,
        bar width=8.5pt,
        ylabel = {Accuracy},
        symbolic x coords={Adload,
				      Agent,
				      Alureon,
				      BHO,
				      CeeInject,
				      Cycbot.G,
				      DelfInject,
				      FakeRean,
				      Hotbar,
				      Lolyda.BF,
				      Obfuscator,
				      OnLineGames,
				      Rbot,
				      Renos,
				      Startpage,
				      Vobfus,
				      Vundo,
				      Winwebsec,
				      Zbot,
				      Zeroaccess},
	y tick label style={
    		/pgf/number format/.cd,
   		fixed,
   		fixed zerofill,
    		precision=0},
        xtick = data,
        x tick label style={
        		rotate=60,
		font=\footnotesize,
		anchor=north east,
		inner sep=0mm},
        nodes near coords,
        every node near coord/.append style={rotate=90, anchor=west, font=\footnotesize},
        enlarge x limits=0.03,
        legend cell align=left,
        legend style={
                at={(0.05,0)},
                anchor=south,
                column sep=1ex
        }
    ]
\addplot[fill=blue,opacity=1.00] 
coordinates {
(Adload,98.05)
(Agent,90.65)
(Alureon,86.74)
(BHO,98.45)
(CeeInject,93.30)
(Cycbot.G,97.80)
(DelfInject,87.65)
(FakeRean,87.94)
(Hotbar,98.70)
(Lolyda.BF,95.70)
(Obfuscator,90.95)
(OnLineGames,95.15)
(Rbot,91.00)
(Renos,91.75)
(Startpage,93.60)
(Vobfus,97.40)
(Vundo,93.35)
(Winwebsec,97.05)
(Zbot,95.05)
(Zeroaccess,99.10)
};
\end{axis}
\end{tikzpicture}
\caption{MLP individual family results}\label{fig:ind_MLP} 
\end{figure}
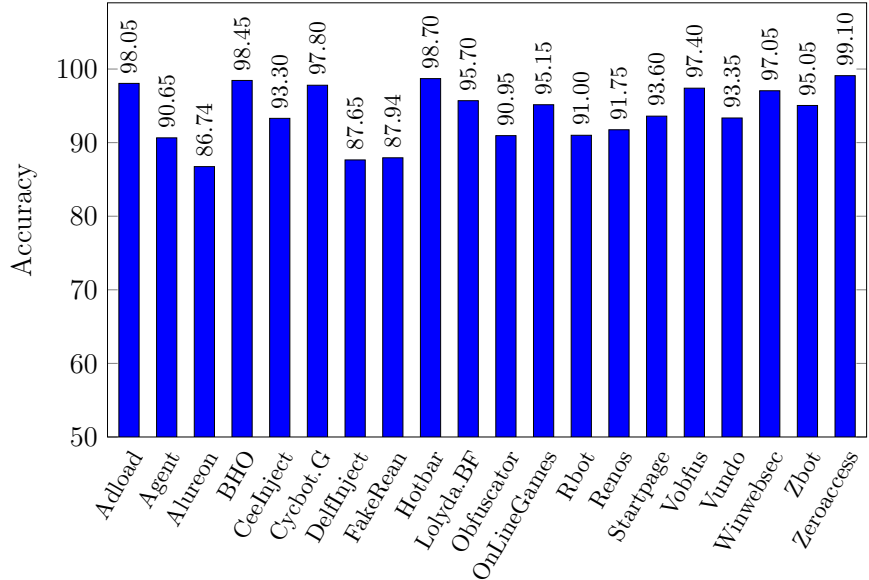

As a final test case, we consider MLP, with the individual family results
given in Figure~\ref{fig:ind_MLP}. For MLP. the average balanced accuracy at 
level~1 is~93.97\%, which is greater than SVM, but below the other two techniques considered. 
We note that Alureon, DelfInject, and FakeRean have the lowest accuracies, 
while Adload, BHO, Hotbar, and Zeroaccess have accuracies greater than~98\%. 
The range in accuracies is~12.36\%, which indicates substantial variability in the results.    

\subsubsection{Summary of Results for Individual Families}

\begin{figure}[!htb]
\centering
\begin{tikzpicture}[scale=0.9]
    \pgfplotstableread[col sep=comma]{data/data.csv}\csvdata
    \pgfplotstabletranspose\datatransposed{\csvdata} 
    \begin{axis}[
        boxplot/draw direction = y,
        axis x line* = bottom,
        axis y line = left,
        enlarge y limits,
        xtick = {1, 2, 3, 4},
        xticklabel style = {align=center, font=\small, rotate=60},
        xticklabels = {SVM, \kNN, RF, MLP},
        ylabel = {Balanced Accuracy},
        	y tick label style={
    		/pgf/number format/.cd,
   		fixed,
   		fixed zerofill,
    		precision=0},
        ytick = {80,85,90,95,100}
    ]
\addplot+[boxplot, thick, draw=black, fill=orange, mark=none] table[y index=4] {\datatransposed};
\addplot+[boxplot, thick, draw=black, fill=green, mark=none] table[y index=2] {\datatransposed};
\addplot+[boxplot, thick, draw=black, fill=red, mark=none] table[y index=1] {\datatransposed};
\addplot+[boxplot, thick, draw=black, fill=yellow, mark=none] table[y index=3] {\datatransposed};
    \end{axis}
\end{tikzpicture}
\caption{Boxplots of individual balanced accuracies}\label{fig:individualBoxplot}
\end{figure}
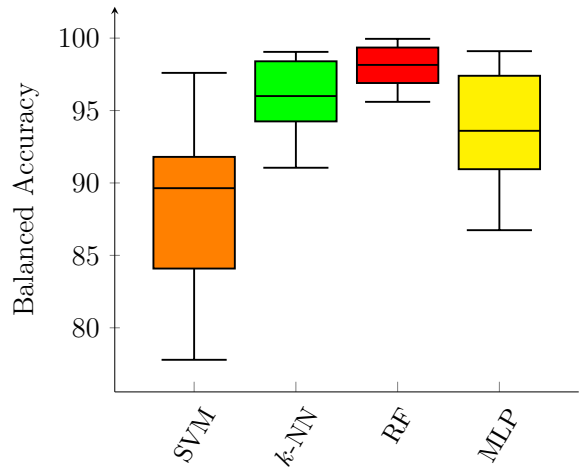

Figure~\ref{fig:individualBoxplot} contains boxplots for the level~1 results for each 
of the four machine learning techniques considered. 
We observe that the SVM box is larger than that for~\kNN, indicating that the accuracies of SVM 
are spread across a wider range, as compared to the accuracies obtained with \kNN. 
The short whisker at the top for \kNN\ box indicates that the data points above the median are closer to 
each other, as compared to the bottom~50\%. 
Note that the box for \kNN\ closely resembles that for RF,
as both have a short whisker at the top and both have shorter boxes compared to MLP and SVM. 
On the other hand,
MLP, like SVM, has a bigger box size, indicating that accuracies are spread over a wider range. 
The length of the whiskers indicate tail length. The SVM long whiskers indicate that the accuracies 
are heavy-tailed, while the \kNN\ data indicate that it has a light-tailed population. 

As noted above, SVM and MLP are closely related techniques, as are 
the \kNN\ and RF algorithms.
Interestingly, the boxplots in Figure~\ref{fig:individualBoxplot} indicate that 
the level~1 results for these pairs both have much in common.

\begin{figure}[!htb]
\centering
    \begin{tikzpicture}[scale=0.8]
    \begin{axis}[width=0.75\textwidth,height=0.675\textwidth,xmin=-0.02,xmax=1.0,
                       ymin=0.0,ymax=1.02,legend pos=south east,grid=both,legend cell align={left},
                       xlabel={False Positive Rate},ylabel={True Positive Rate},
                       xtick={0.00,0.20,0.40,0.60,0.80,1.00},ytick={0.00,0.20,0.40,0.60,0.80,1.00}] 
         \pgfplotstableread{data/fig12_svm.txt}\mydata;
         \addplot[color=orange,ultra thick] 
                 table 
                 [
                   x expr =\thisrowno{0}, 
                   y expr =\thisrowno{1}
                 ] {\mydata};
                 \addlegendentry{SVM ($\AUC = 0.86$)}
         \pgfplotstableread{data/fig12_knn.txt}\mydata;
         \addplot[color=green,ultra thick] 
                 table 
                 [
                   x expr =\thisrowno{0}, 
                   y expr =\thisrowno{1}
                 ] {\mydata};
                 \addlegendentry{\kNN\ ($\AUC = 0.94$)}
         \pgfplotstableread{data/fig12_rf.txt}\mydata;
         \addplot[color=red,ultra thick] 
                 table 
                 [
                   x expr =\thisrowno{0}, 
                   y expr =\thisrowno{1}
                 ] {\mydata};
                 \addlegendentry{RF ($\AUC = 0.99$)}
         \pgfplotstableread{data/fig12_mlp.txt}\mydata;
         \addplot[color=yellow,ultra thick] 
                 table 
                 [
                   x expr =\thisrowno{0}, 
                   y expr =\thisrowno{1}
                 ] {\mydata};
                 \addlegendentry{MLP ($\AUC = 0.90$)}
    \end{axis}
    \end{tikzpicture}
\caption{ROC curves for DelfInject}\label{fig:rocDelfInject}
\end{figure}
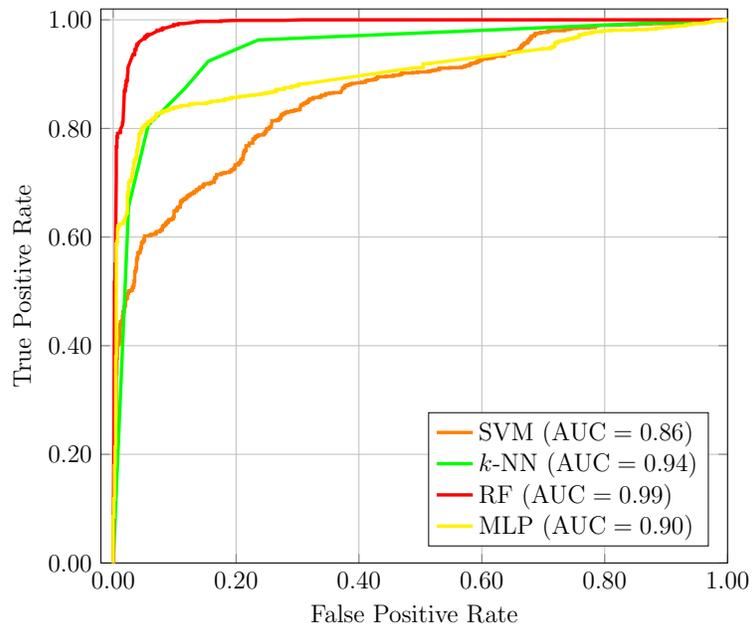

The DelfInject family is among the most difficult families to detect for each of the four ML
techniques considered. Figure~\ref{fig:rocDelfInject} shows the ROC curves for this family. 
ROC curves graphically illustrate the relationship between the false positive rate and the 
true positive rate as the threshold passes through the range of values. 
The area under the ROC curve (AUC) ranges from~0 to~1, and can be interpreted as
the probability that a randomly selected positive instance scores higher than a 
randomly selected negative instance~\cite{stamp2017introduction}. 
The AUC values range from~0.99 to~0.86,
with RF performing best, and SVM performing the worst.

\subsection{Multifamily Models}

We now consider the effect of modeling multiple families. 
For each level---and for each of the four machine learning techniques under 
consideration---we graph the average balanced accuracies, as well as the
the lowest, and the highest accuracies. As with the individual family models
in the previous section, all experimental results
reported in this section are based on bigrams.

\subsubsection{Support Vector Machines}

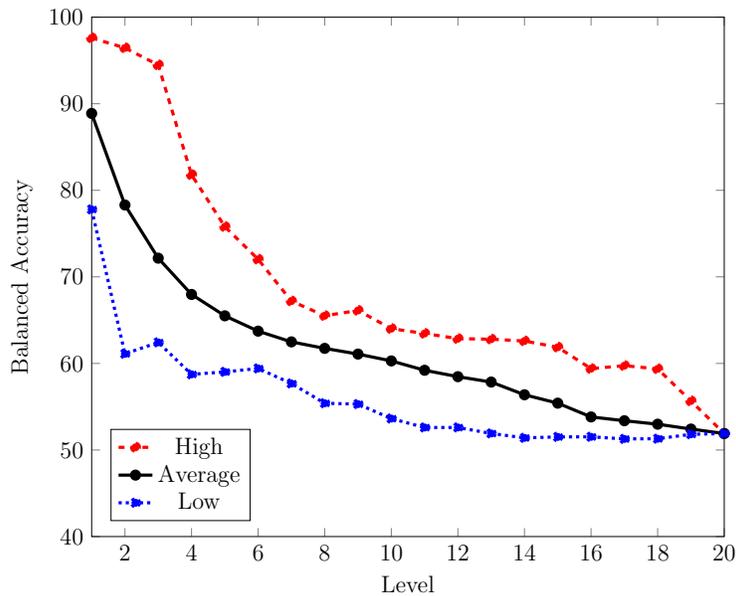
\begin{figure}[!htb]
\centering
\begin{tikzpicture}[scale=0.75]
\begin{axis}[width=0.80\textwidth,
		   height=0.675\textwidth,
	 	   x tick label style={
    		 	/pgf/number format/.cd,
   			fixed,
   			fixed zerofill,
    			precision=0},
	 	   y tick label style={
    		 	/pgf/number format/.cd,
   			fixed,
   			fixed zerofill,
    			precision=0},
                    xmin=1,xmax=20,
                    ymin=40.0,ymax=100.0,
                    legend pos=south west,
                    xlabel={Level},
                    ylabel={Balanced Accuracy}] 
\addplot[color=red,ultra thick,mark=*,mark size=2.0,dashed] coordinates {
(1,97.60)
(2,96.45)
(3,94.48)
(4,81.79)
(5,75.80)
(6,72.03)
(7,67.20)
(8,65.51)
(9,66.12)
(10,64.02)
(11,63.46)
(12,62.86)
(13,62.80)
(14,62.59)
(15,61.87)
(16,59.40)
(17,59.74)
(18,59.37)
(19,55.66)
(20,51.90)
};
\addplot[color=black,ultra thick,mark=*,mark size=2.0] coordinates {
(1,88.88) 
(2,78.30) 
(3,72.16) 
(4,67.97) 
(5,65.50) 
(6,63.73) 
(7,62.49) 
(8,61.74) 
(9,61.08)
(10,60.29) 
(11,59.21) 
(12,58.47) 
(13,57.85) 
(14,56.38) 
(15,55.41) 
(16,53.82) 
(17,53.38) 
(18,52.98) 
(19,52.42) 
(20,51.90)
};
\addplot[color=blue,ultra thick,mark=*,mark size=2.0,dotted] coordinates {
(1,77.79)
(2,61.11)
(3,62.43)
(4,58.74)
(5,59.02)
(6,59.42)
(7,57.68)
(8,55.40)
(9,55.30)
(10,53.63)
(11,52.60)
(12,52.60)
(13,51.91)
(14,51.40)
(15,51.51)
(16,51.52)
(17,51.28)
(18,51.30)
(19,51.81)
(20,51.90)
};
\legend{High,Average,Low}
\end{axis}
\end{tikzpicture}
\caption{SVM multifamily results (high, average, and low)}\label{fig:svmgraph}
\end{figure}

Figure~\ref{fig:svmgraph} contains graphs of the high, average, and low 
balanced accuracies of multifamily 
models using SVM. As mentioned in the previous section, at level~1, 
the average balanced accuracy is~88.88\%. For models with pairs of 
malware families combined, the average balanced accuracy drops to~78.30\%,
and at level~20, the accuracy drops to essentially a coin flip, at~51.90\%. 

From level~1 to level~5, the average accuracy declines drastically, with the largest 
drop from level~1 to level~2. These results indicate that the SVM does not generalize 
well to the multifamily case. 

\subsubsection{$k$-Nearest Neighbors (\kNN)}

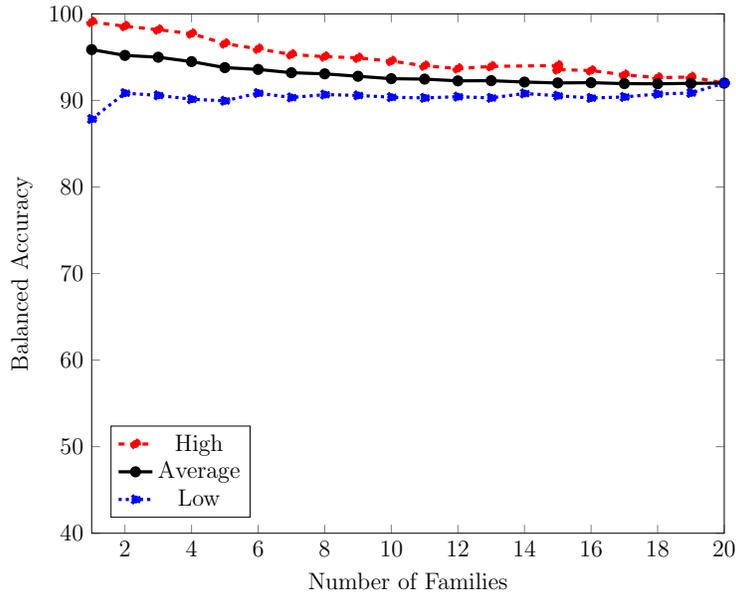
\begin{figure}[!htb]
\centering
\begin{tikzpicture}[scale=0.75]
\begin{axis}[width=0.80\textwidth,
		   height=0.675\textwidth,
	 	   x tick label style={
    		 	/pgf/number format/.cd,
   			fixed,
   			fixed zerofill,
    			precision=0},
	 	   y tick label style={
    		 	/pgf/number format/.cd,
   			fixed,
   			fixed zerofill,
    			precision=0},
                    xmin=1,xmax=20,
                    ymin=40.0,ymax=100.0,
                    legend pos=south west,
                    xlabel={Number of Families},
                    ylabel={Balanced Accuracy}] 
\addplot[color=red,ultra thick,mark=*,mark size=2.0,dashed] coordinates {
(1,99.05)
(2,98.57)
(3,98.17)
(4,97.71)
(5,96.56)
(6,95.96)
(7,95.30)
(8,95.06)
(9,94.91)
(10,94.54)
(11,94.00)
(12,93.68)
(13,93.93)
(15,94.03)
(15,93.56)
(16,93.45)
(17,92.96)
(18,92.63)
(19,92.70)
(20,92.00)
};
\addplot[color=black,ultra thick,mark=*,mark size=2.0] coordinates {
(1,95.87)
(2,95.20)
(3,94.99)
(4,94.48)
(5,93.79)
(6,93.58)
(7,93.20)
(8,93.07)
(9,92.79)
(10,92.51)
(11,92.46)
(12,92.26)
(13,92.28)
(14,92.12)
(15,92.02)
(16,92.05)
(17,91.94)
(18,91.92)
(19,91.97)
(20,92.00)
};
\addplot[color=blue,ultra thick,mark=*,mark size=2.0,dotted] coordinates {
(1,87.84)
(2,90.84)
(3,90.58)
(4,90.14)
(5,89.95)
(6,90.84)
(7,90.34)
(8,90.67)
(9,90.58)
(10,90.37)
(11,90.29)
(12,90.43)
(13,90.29)
(14,90.80)
(15,90.54)
(16,90.28)
(17,90.40)
(18,90.74)
(19,90.86)
(20,92.00)
};
\legend{High,Average,Low}
\end{axis}
\end{tikzpicture}
\caption{\kNN\ multifamily results (high, average, and low)}\label{fig:knngraph}
\end{figure}

Figure~\ref{fig:knngraph} shows a line graph representing the average, the lowest, and the highest 
balanced accuracies at each level for~\kNN. The overall trend shows that as more
families are added, the performance decreases. However, the accuracy is 
over~90\%\ even when modeling all~20 families. Note that this is in stark contrast to
the SVM results presented in the previous section, where the accuracy
dropped off dramatically as families were added to the malware class.

\subsubsection{Random Forest}

\begin{figure}[!htb]
\centering
\begin{tikzpicture}[scale=0.75]
\begin{axis}[width=0.80\textwidth,
		   height=0.675\textwidth,
	 	   x tick label style={
    		 	/pgf/number format/.cd,
   			fixed,
   			fixed zerofill,
    			precision=0},
	 	   y tick label style={
    		 	/pgf/number format/.cd,
   			fixed,
   			fixed zerofill,
    			precision=0},
                    xmin=1,xmax=20,
                    ymin=50.0,ymax=100.0,
                    legend pos=south west,
                    xlabel={Number of Malware Families},
                    ylabel={Average Balanced Accuracy}] 
\addplot[color=red,ultra thick,mark=*,mark size=2.0] coordinates {
(1,98.23)
(5,96.77)
(10,94.81)
(15,93.35)
(20,92.87)
};
\addplot[color=blue,ultra thick,mark=*,mark size=2.0] coordinates {
(1,97.46)
(5,92.34)
(10,86.10)
(15,82.90)
(20,82.53)
};
\addplot[color=green,ultra thick,mark=*,mark size=2.0] coordinates {
(1,92.85)
(5,70.01)
(10,56.78)
(15,53.98)
(20,50.00)
};
\legend{depth~10,depth~5,depth~2}
\end{axis}
\end{tikzpicture}
\caption{Random forest results for various depth values}\label{fig:RFbalancedAccuracydepth}
\end{figure}
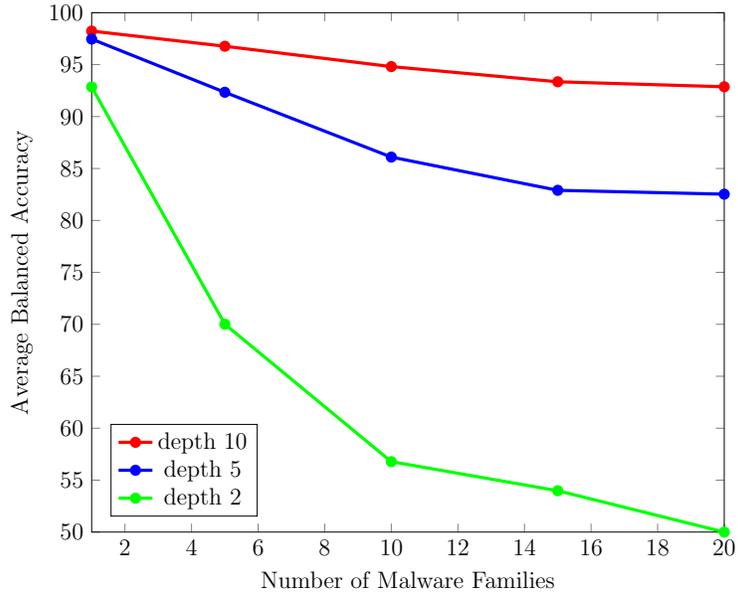

For our RF multifamily experiments, we first experiment with the depth parameter at 
levels~1, 5, 10, 15, and~20. Figure~\ref{fig:RFbalancedAccuracydepth} 
gives the results of these experiments.  We observe that at level~1, a depth of~10
gives the best results, and this persists at higher levels. Thus, for all of the experiments
that we report here, we have used a depth of~10.

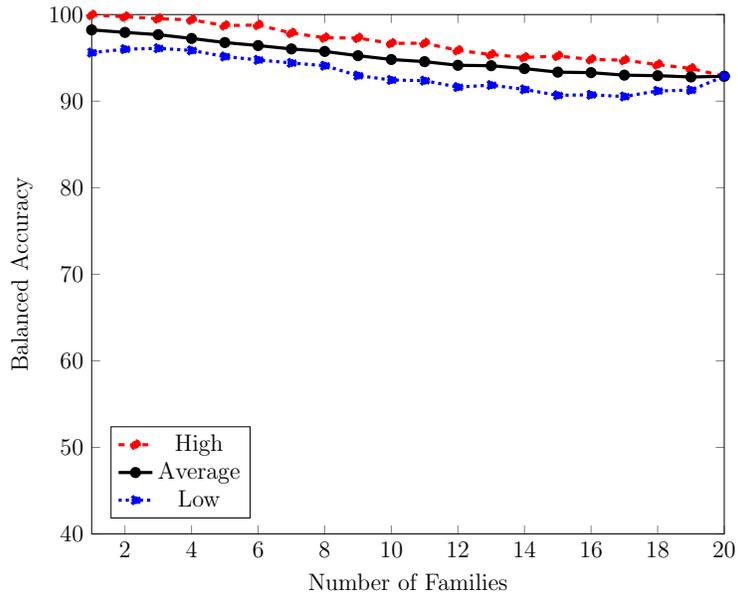
\begin{figure}[!htb]
\centering
\begin{tikzpicture}[scale=0.75]
\begin{axis}[width=0.80\textwidth,
		   height=0.675\textwidth,
	 	   x tick label style={
    		 	/pgf/number format/.cd,
   			fixed,
   			fixed zerofill,
    			precision=0},
	 	   y tick label style={
    		 	/pgf/number format/.cd,
   			fixed,
   			fixed zerofill,
    			precision=0},
                    xmin=1,xmax=20,
                    ymin=40.0,ymax=100.0,
                    legend pos=south west,
                    xlabel={Number of Families},
                    ylabel={Balanced Accuracy}] 
\addplot[color=red,ultra thick,mark=*,mark size=2.0,dashed] coordinates {
(1,99.95)
(2,99.75)
(3,99.52)
(4,99.39)
(5,98.73)
(6,98.79)
(7,97.87)
(8,97.32)
(9,97.29)
(10,96.67)
(11,96.70)
(12,95.86)
(13,95.38)
(14,95.05)
(15,95.21)
(16,94.82)
(17,94.73)
(18,94.22)
(19,93.78)
(20,92.87)
};
\addplot[color=black,ultra thick,mark=*,mark size=2.0] coordinates {
(1,98.23)
(2,97.94)
(3,97.68)
(4,97.24)
(5,96.77)
(6,96.42)
(7,96.03)
(8,95.74)
(9,95.25)
(10,94.81)
(11,94.56)
(12,94.14)
(13,94.09)
(14,93.76)
(15,93.35)
(16,93.29)
(17,93.00)
(18,92.94)
(19,92.79)
(20,92.87)
};
\addplot[color=blue,ultra thick,mark=*,mark size=2.0,dotted] coordinates {
(1,95.60)
(2,96.00)
(3,96.10)
(4,95.86)
(5,95.15)
(6,94.76)
(7,94.41)
(8,94.09)
(9,92.95)
(10,92.44)
(11,92.36)
(12,91.62)
(13,91.85)
(14,91.35)
(15,90.67)
(16,90.73)
(17,90.53)
(18,91.18)
(19,91.27)
(20,92.87)
};
\legend{High,Average,Low}
\end{axis}
\end{tikzpicture}
\caption{RF multifamily results (high, average, and low)}\label{fig:rfgraph}
\end{figure}

Figure~\ref{fig:rfgraph} shows three lines representing the average, the lowest, 
and the highest balanced accuracies for multifamily models at each level. Again,
these results were obtained using RF with a depth of~10. 
The average balanced accuracy for classifying a specific family as malware or benign is~98.23\%. 
When two families are combined, the average balanced accuracy drops slightly to~97.94\%,
and when all~20 families are combined in the malware class, 
the accuracy is still a respectable~92.87\%.

\subsubsection{Multilayer Perceptron}

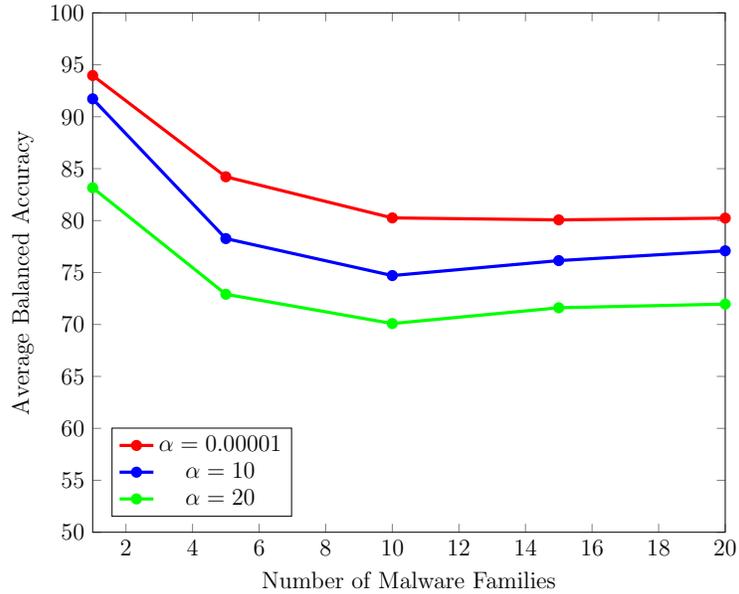
\begin{figure}[!htb]
\centering
\begin{tikzpicture}[scale=0.75]
\begin{axis}[width=0.80\textwidth,
		   height=0.675\textwidth,
	 	   x tick label style={
    		 	/pgf/number format/.cd,
   			fixed,
   			fixed zerofill,
    			precision=0},
	 	   y tick label style={
    		 	/pgf/number format/.cd,
   			fixed,
   			fixed zerofill,
    			precision=0},
                    xmin=1,xmax=20,
                    ymin=50.0,ymax=100.0,
                    legend pos=south west,
                    xlabel={Number of Malware Families},
                    ylabel={Average Balanced Accuracy}] 
\addplot[color=red,ultra thick,mark=*,mark size=2.0] coordinates {
(1,93.97)
(5, 84.22)
(10, 80.26)
(15, 80.07)
(20, 80.24)
};
\addplot[color=blue,ultra thick,mark=*,mark size=2.0] coordinates {
(1,91.72)
(5,78.27)
(10,74.71)
(15,76.15)
(20,77.09)
};
\addplot[color=green,ultra thick,mark=*,mark size=2.0] coordinates {
(1,83.16)
(5,72.91)
(10,70.09)
(15,71.61)
(20,71.96)
};
\legend{$\alpha=0.00001$,$\alpha=10$,$\alpha=20$}
\end{axis}
\end{tikzpicture}
\caption{MLP balanced accuracies for various~$\alpha$ values}\label{fig:MLPalpha}
\end{figure}

In an MLP, the regularization (or penalty) parameter~$\alpha$ is used to reduce
overfitting~\cite{scikitlearn}. In Figure~\ref{fig:MLPalpha}, we give results for MLP 
experiments as the parameter~$\alpha$ varies. 
These results clearly indicate that a model with~$\alpha=0.00001$ 
performed the best from among those tested, 
and hence we use this value in all subsequent experiments.

\begin{figure}[!htb]
\centering
\begin{tikzpicture}[scale=0.75]
\begin{axis}[width=0.80\textwidth,
		   height=0.675\textwidth,
	 	   x tick label style={
    		 	/pgf/number format/.cd,
   			fixed,
   			fixed zerofill,
    			precision=0},
	 	   y tick label style={
    		 	/pgf/number format/.cd,
   			fixed,
   			fixed zerofill,
    			precision=0},
                    xmin=1,xmax=20,
                    ymin=40.0,ymax=100.0,
                    legend pos=south west,
                    xlabel={Number of Families},
                    ylabel={Balanced Accuracy}] 
\addplot[color=red,ultra thick,mark=*,mark size=2.0,dashed] coordinates {
(1,99.10)
(2,98.15)
(3,97.90)
(4,94.01)
(5,93.70)
(6,93.26)
(7,92.49)
(8,91.20)
(9,92.00)
(10,92.25)
(11,91.94)
(12,89.64)
(13,90.43)
(14,87.96)
(15,89.11)
(16,86.24)
(17,85.34)
(18,84.40)
(19,84.30)
(20,80.24)
};
\addplot[color=black,ultra thick,mark=*,mark size=2.0] coordinates {
(1,93.97)
(2,90.45)
(3,89.10)
(4,87.29)
(5,84.22)
(6,83.63)
(7,81.82)
(8,81.47)
(9,81.06)
(10,80.26)
(11,80.59)
(12,80.27)
(13,79.92)
(14,80.36)
(15,80.07)
(16,80.37)
(17,80.14)
(18,79.82)
(19,79.45)
(20,80.24)
};
\addplot[color=blue,ultra thick,mark=*,mark size=2.0,dotted] coordinates {
(1,86.73)
(2,75.37)
(3,77.82)
(4,73.59)
(5,70.01)
(6,60.02)
(7,70.24)
(8,58.27)
(9,63.62)
(10,63.26)
(11,63.31)
(12,66.51)
(13,65.37)
(14,61.93)
(15,61.89)
(16,62.50)
(17,66.17)
(18,66.37)
(19,68.37)
(20,80.24)
};
\legend{High,Average,Low}
\end{axis}
\end{tikzpicture}
\caption{MLP multifamily results (high, average, and low)}\label{fig:nngraph}
\end{figure}
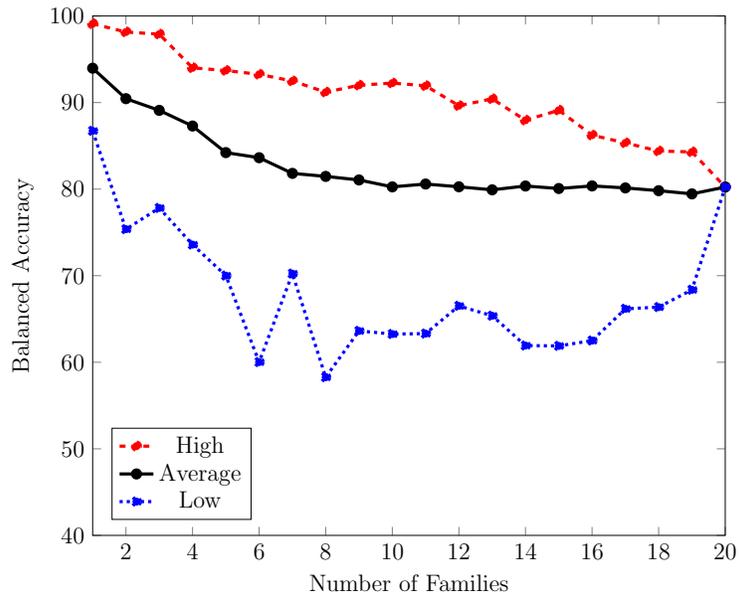

Figure~\ref{fig:nngraph} gives the lowest, highest, and average accuracies at each level in
our MLP experiments. As expected, for the average case, the trend is downward. In comparison
to the other ML techniques considered, we see that MLP performs somewhat better
than SVM, but with a larger variability, particularly with respect to the minimum accuracy.

\subsection{Summary of Bigram Results}

\begin{figure}[!htb]
\centering
\begin{tikzpicture}[scale=0.75]
\begin{axis}[width=0.80\textwidth,
		   height=0.675\textwidth,
	 	   x tick label style={
    		 	/pgf/number format/.cd,
   			fixed,
   			fixed zerofill,
    			precision=0},
	 	   y tick label style={
    		 	/pgf/number format/.cd,
   			fixed,
   			fixed zerofill,
    			precision=0},
                    xmin=1,xmax=20,
                    ymin=40.0,ymax=100.0,
                    legend pos=south west,
                    xlabel={Number of Families},
                    ylabel={Balanced Accuracy}] 
\addplot[color=red,ultra thick,mark=*,mark size=2.0] coordinates {
(1,98.23)
(2,97.94)
(3,97.68)
(4,97.24)
(5,96.77)
(6,96.42)
(7,96.03)
(8,95.74)
(9,95.25)
(10,94.81)
(11,94.56)
(12,94.14)
(13,94.09)
(14,93.76)
(15,93.35)
(16,93.29)
(17,93.00)
(18,92.94)
(19,92.79)
(20,92.87)
};
\addplot[color=green,ultra thick,mark=*,mark size=2.0] coordinates {
(1,95.87)
(2,95.20)
(3,94.99)
(4,94.48)
(5,93.79)
(6,93.58)
(7,93.20)
(8,93.07)
(9,92.79)
(10,92.51)
(11,92.46)
(12,92.26)
(13,92.28)
(14,92.12)
(15,92.02)
(16,92.05)
(17,91.94)
(18,91.92)
(19,91.97)
(20,92.00)
};
\addplot[color=yellow,ultra thick,mark=*,mark size=2.0] coordinates {
(1,93.97)
(2,90.45)
(3,89.10)
(4,87.29)
(5,84.22)
(6,83.63)
(7,81.82)
(8,81.47)
(9,81.06)
(10,80.26)
(11,80.59)
(12,80.27)
(13,79.92)
(14,80.36)
(15,80.07)
(16,80.37)
(17,80.14)
(18,79.82)
(19,79.45)
(20,80.24)
};
\addplot[color=orange,ultra thick,mark=*,mark size=2.0] coordinates {
(1,88.88) 
(2,78.30) 
(3,72.16) 
(4,67.97) 
(5,65.50) 
(6,63.73) 
(7,62.49) 
(8,61.74) 
(9,61.08)
(10,60.29) 
(11,59.21) 
(12,58.47) 
(13,57.85) 
(14,56.38) 
(15,55.41) 
(16,53.82) 
(17,53.38) 
(18,52.98) 
(19,52.42) 
(20,51.90)
};
\legend{RF,\kNN,MLP,SVM}
\end{axis}
\end{tikzpicture}
\caption{Average multifamily balanced accuracies}\label{fig:balancedAccuracy}
\end{figure}
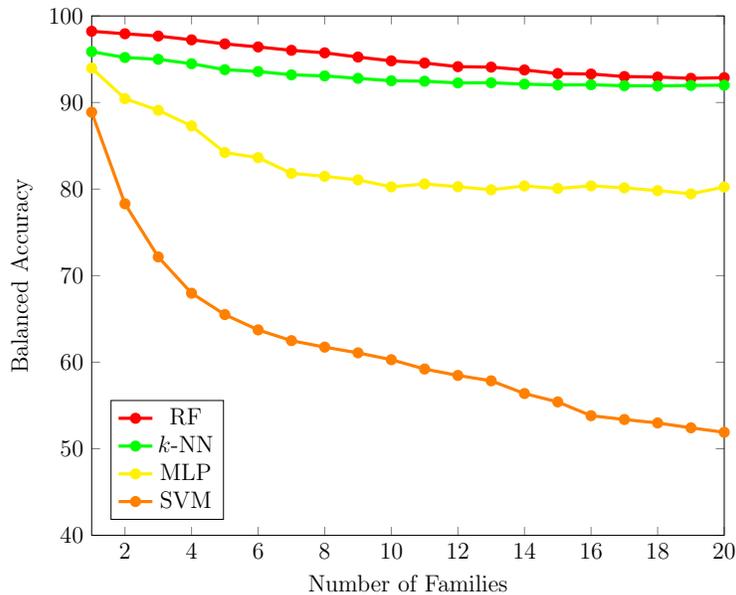

Figure~\ref{fig:balancedAccuracy} summarizes the performance of all the 
four ML techniques considered above. This graph shows how the average 
balanced accuracy changes from single-family binary classification to 
twenty-family binary classification for SVM, \kNN, random forest, and MLP. 
Qualitatively, we observe that for each ML technique, 
the accuracy trends downward as more families are added to the malware class.
However, quantitatively, the differences between the four techniques are dramatic.
We see that SVM performs relatively poorly, while the two neighborhood-based techniques
(\kNN\ and RF) are surprisingly robust as more families are added to
the malware class. The MLP experiments are in between the neighborhood-based
techniques and SVM.

\begin{figure}[!htb]
\centering
\begin{tabular}{cccc}
\begin{tikzpicture}
    \pgfplotstableread[col sep=comma]{data/data_19.csv}\csvdata
    \pgfplotstabletranspose\datatransposed{\csvdata} 
    \begin{axis}[width=0.225\textwidth,height=0.4\textwidth,
        boxplot/draw direction = y,
        axis x line* = bottom,
        axis y line = left,
        enlarge y limits,
        xtick = {1},
        xticklabel style = {align=center, font=\small},
        xticklabels = {},
        ylabel = {Balanced Accuracy},
        	y tick label style={
		font=\small,
    		/pgf/number format/.cd,
   		fixed,
   		fixed zerofill,
    		precision=0},
        ytick = {52,53,54,55,56}
    ]
\addplot+[boxplot, thick, draw=black, fill=orange, mark=none] table[y index=4] {\datatransposed};
    \end{axis}
\end{tikzpicture}
&
\begin{tikzpicture}
    \pgfplotstableread[col sep=comma]{data/data_19.csv}\csvdata
    \pgfplotstabletranspose\datatransposed{\csvdata} 
    \begin{axis}[width=0.225\textwidth,height=0.4\textwidth,
        boxplot/draw direction = y,
        axis x line* = bottom,
        axis y line = left,
        enlarge y limits,
        xtick = {1},
        xticklabel style = {align=center, font=\small}, 
        xticklabels = {},
        	y tick label style={
		font=\small,
    		/pgf/number format/.cd,
   		fixed,
   		fixed zerofill,
    		precision=1},
        ytick = {91.0,91.5,92.0,92.5,93.0}
    ]
\addplot+[boxplot, thick, draw=black, fill=green, mark=none] table[y index=2] {\datatransposed};
    \end{axis}
\end{tikzpicture}
&
\begin{tikzpicture}
    \pgfplotstableread[col sep=comma]{data/data_19.csv}\csvdata
    \pgfplotstabletranspose\datatransposed{\csvdata} 
    \begin{axis}[width=0.225\textwidth,height=0.4\textwidth,
        boxplot/draw direction = y,
        axis x line* = bottom,
        axis y line = left,
        enlarge y limits,
        xtick = {1},
        xticklabel style = {align=center, font=\small},
        xticklabels = {},
        	y tick label style={
		font=\small,
    		/pgf/number format/.cd,
   		fixed,
   		fixed zerofill,
    		precision=1},
        ytick = {91.5,92.0,92.5,93.0,93.5,94.0}
    ]
\addplot+[boxplot, thick, draw=black, fill=red, mark=none] table[y index=1] {\datatransposed};
    \end{axis}
\end{tikzpicture}
&
\begin{tikzpicture}
    \pgfplotstableread[col sep=comma]{data/data_19.csv}\csvdata
    \pgfplotstabletranspose\datatransposed{\csvdata} 
    \begin{axis}[width=0.225\textwidth,height=0.4\textwidth,
        boxplot/draw direction = y,
        axis x line* = bottom,
        axis y line = left,
        enlarge y limits,
        xtick = {1},
        xticklabel style = {align=center, font=\small}, 
        xticklabels = {},
        	y tick label style={
		font=\small,
    		/pgf/number format/.cd,
   		fixed,
   		fixed zerofill,
    		precision=0},
        ytick = {68,72,76,80,84}
    ]
\addplot+[boxplot, thick, draw=black, fill=yellow, mark=none] table[y index=3] {\datatransposed};
    \end{axis}
\end{tikzpicture}
\\
\hspace*{0.375in} (a) SVM & 
\hspace*{0.3in} (b) \kNN & 
\hspace*{0.25in} (c) RF & 
\hspace*{0.25in} (d) MLP
\end{tabular}
\caption{Boxplots of balanced accuracy with~19 families}\label{fig:19_boxplot}
\end{figure}
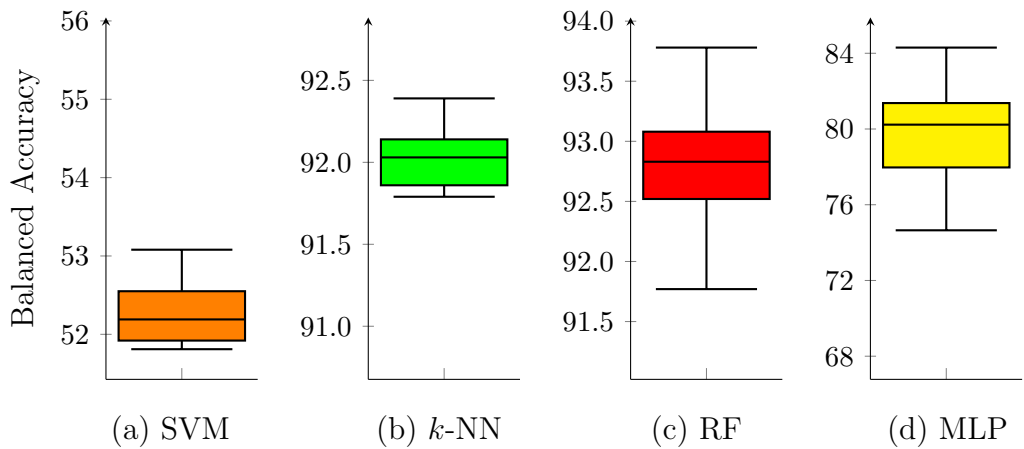

In Figure~\ref{fig:19_boxplot}, we summarize the variability 
of the four ML techniques at level~19,
which is the highest level for which such plots are applicable.
Be sure to note the different scales on the boxplots in Figure~\ref{fig:19_boxplot}~(a)
through~(d). These results serve to further emphasize the superiority 
of the neighborhood-based techniques, as well as to highlight some of the performance
differences between the various techniques. While RF performs slightly better than \kNN\ on average,
we see that \kNN\ has the smallest variance at this level, and hence achieves more consistent results,
with nearly the same average accuracy. The SVM is more consistent than the MLP,
but its performance is at such a low accuracy that it is of no use as a classifier.

As mentioned above, an MLP is comparable to an SVM where
the kernel function is determined from the training data, rather than being
specified by the user. Thus, given sufficient training data, we would generally expect
an MLP to outperform an SVM, and indeed we do find this to be the case
in our ML experiments.

Also as mentioned above, we would generally expect \kNN\ and RF to perform
somewhat similarly, since both are neighborhood-based techniques.
When viewed from this perspective, it is not surprising that these two techniques
yield comparable results. However, it is surprising that the two neighborhood-based
ML techniques perform so well, with both attaining at least~92\%\ accuracy
over a malware dataset consisting of~20 diverse families.

It is not entirely clear why neighborhood-based techniques such as \kNN\ and random forest 
would be so robust in the face of such diverse data. But, we can say
that the decision boundaries generated in our SVM and MLP experiments 
are clearly inferior to those based simply on proximity (by some measure)
to known malware samples. It does make some intuitive sense that the best
predictor of malware is ``closeness'' to other known malware samples.

\subsection{Multifamily Models with 4-grams and 6-grams}\label{sect:4-6-grams}

All of the results above are based on bigram features.
In this section, we give balanced accuracy results analogous to
those in Figure~\ref{fig:balancedAccuracy} for 
experiments based on 4-gram and 6-gram features.

\begin{figure}[!htb]
\centering
\begin{tabular}{cc}
\begin{tikzpicture}[scale=0.5]
\begin{axis}[width=0.80\textwidth,
		   height=0.675\textwidth,
	 	   x tick label style={
    		 	/pgf/number format/.cd,
   			fixed,
   			fixed zerofill,
    			precision=0},
	 	   y tick label style={
    		 	/pgf/number format/.cd,
   			fixed,
   			fixed zerofill,
    			precision=0},
                    xmin=1,xmax=20,
                    ymin=40.0,ymax=100.0,
                    legend pos=south west,
                    xlabel={Number of Families},
                    ylabel={Balanced Accuracy}] 
\addplot[color=red,ultra thick,mark=*,mark size=2.0] coordinates {
(1,95.47)
(2,94.71)
(3,93.38)
(4,93.49)
(5,93.38)
(6,92.72)
(7,92.44)
(8,92.20)
(9,91.83)
(10,91.17)
(11,91.18)
(12,90.39)
(13,91.09)
(14,90.48)
(15,89.99)
(16,89.94)
(17,90.39)
(18,90.25)
(19,90.19)
(20,90.20)
};
\addplot[color=green,ultra thick,mark=*,mark size=2.0] coordinates {
(1,94.03)
(2,92.96)
(3,92.01)
(4,92.31)
(5,92.20)
(6,91.81)
(7,91.63)
(8,91.62)
(9,91.35)
(10,91.16)
(11,91.09)
(12,90.65)
(13,91.06)
(14,90.76)
(15,90.68)
(16,90.63)
(17,90.81)
(18,90.66)
(19,90.67)
(20,90.73)
};
\addplot[color=yellow,ultra thick,mark=*,mark size=2.0] coordinates {
(1,89.52)
(2,83.92)
(3,80.56)
(4,79.94)
(5,81.49)
(6,81.10)
(7,79.96)
(8,81.96)
(9,80.87)
(10,79.97)
(11,80.55)
(12,80.03)
(13,81.85)
(14,79.97)
(15,79.38)
(16,80.02)
(17,81.34)
(18,80.68)
(19,80.97)
(20,81.15)
};
\addplot[color=orange,ultra thick,mark=*,mark size=2.0] coordinates {
(1,85.95)
(2,71.92)
(3,67.35)
(4,66.38)
(5,65.88)
(6,65.23)
(7,63.94)
(8,63.42)
(9,62.32)
(10,61.69)
(11,61.16)
(12,60.64)
(13,60.47)
(14,59.96)
(15,59.53)
(16,59.02)
(17,58.62)
(18,58.35)
(19,58.16)
(20,57.87)
};
\legend{RF,\kNN,MLP,SVM}
\end{axis}
\end{tikzpicture}
&
\begin{tikzpicture}[scale=0.5]
\begin{axis}[width=0.80\textwidth,
		   height=0.675\textwidth,
	 	   x tick label style={
    		 	/pgf/number format/.cd,
   			fixed,
   			fixed zerofill,
    			precision=0},
	 	   y tick label style={
    		 	/pgf/number format/.cd,
   			fixed,
   			fixed zerofill,
    			precision=0},
                    xmin=1,xmax=20,
                    ymin=40.0,ymax=100.0,
                    legend pos=south west,
                    xlabel={Number of Families},
                    ylabel={Balanced Accuracy}] 
\addplot[color=red,ultra thick,mark=*,mark size=2.0] coordinates {
(1,94.79)
(2,94.49)
(3,94.64)
(4,93.85)
(5,92.66)
(6,92.79)
(7,92.52)
(8,92.49)
(9,92.26)
(10,92.06)
(11,91.42)
(12,91.53)
(13,91.45)
(14,91.06)
(15,90.70)
(16,90.60)
(17,90.70)
(18,90.68)
(19,90.43)
(20,90.07)
};
\addplot[color=green,ultra thick,mark=*,mark size=2.0] coordinates {
(1,91.11)
(2,92.75)
(3,92.96)
(4,92.45)
(5,91.61)
(6,91.85)
(7,91.72)
(8,91.82)
(9,91.78)
(10,91.76)
(11,91.50)
(12,91.57)
(13,91.40)
(14,91.30)
(15,91.22)
(16,91.13)
(17,91.11)
(18,90.99)
(19,91.05)
(20,90.72)
};
\addplot[color=yellow,ultra thick,mark=*,mark size=2.0] coordinates {
(1,85.19)
(2,84.65)
(3,84.30)
(4,83.47)
(5,81.31)
(6,83.13)
(7,83.38)
(8,83.79)
(9,82.65)
(10,83.81)
(11,83.19)
(12,84.09)
(13,83.36)
(14,83.04)
(15,82.69)
(16,82.04)
(17,82.52)
(18,82.50)
(19,82.74)
(20,83.43)
};
\addplot[color=orange,ultra thick,mark=*,mark size=2.0] coordinates {
(1,83.50)
(2,68.48)
(3,61.79)
(4,59.00)
(5,56.09)
(6,55.76)
(7,54.76)
(8,54.76)
(9,54.66)
(10,54.35)
(11,54.50)
(12,54.04)
(13,53.79)
(14,53.45)
(15,53.61)
(16,53.54)
(17,53.53)
(18,53.46)
(19,53.39)
(20,53.33)
};
\legend{RF,\kNN,MLP,SVM}
\end{axis}
\end{tikzpicture}
\\
(a) 4-gram
&
(b) 6-gram
\end{tabular}
\caption{Average multifamily balanced accuracies for higher $n$-grams}
\label{fig:balancedAccuracy_4_6}
\end{figure}
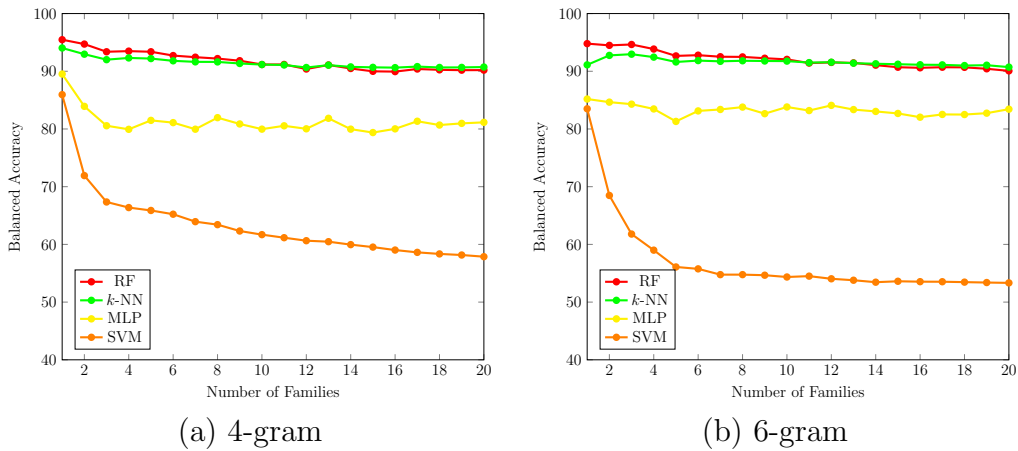

Figure~\ref{fig:balancedAccuracy_4_6}~(a) gives
the accuracies for each of the four machine learning models
under consideration in the case of 4-gram features, whereas
Figure~\ref{fig:balancedAccuracy_4_6}~(b) gives the
corresponding results for experiments based on 6-gram features.

Comparing Figures~\ref{fig:balancedAccuracy_4_6}~(a) and~(b),
we see that qualitatively, the models based on 4-grams
and those based on 6-grams are similar. Comparing the graphs 
in Figure~\ref{fig:balancedAccuracy_4_6}
to the corresponding results in Figure~\ref{fig:balancedAccuracy},
we also observe that the bigram graphs are qualitatively similar
to those for the higher $n$-grams. Quantitatively, the differences
are slight, but the bigrams generally outperform 4-grams,
and 4-grams are marginally better than 6-grams, with the higher $n$-grams
being more competitive (and occasionally superior)
for models involving larger numbers of families.
The only significant advantage for higher $n$-grams is
for the 4-gram SVM model, but in all cases, the SVM model is the weakest
of the four machine learning techniques tested.
Since it is more efficient to collect bigram features, and since we find
no clear advantage to the higher $n$-grams, we conclude
that bigram features are preferred over 4-grams or 6-grams.

\section{Conclusion and Future Work}\label{chap:conclusion}

In this paper, we have  
carefully examined the change in the balanced accuracy as models are trained 
over increasingly diverse malware sets.
We experimented with multiple machine learning techniques
and considered byte $n$-gram features for~$n=2$, $n=4$, and~$n=6$. 

\begin{figure}[!htb]
\centering
\begin{tikzpicture}[scale=0.9, every node/.style={scale=1.0}]
    \begin{axis}[
        width  = 0.7*\textwidth,
        height = 8cm,
        ymin=40.0,ymax=110.0,
        ytick={40,50,60,70,80,90,100},
        major x tick style = transparent,
        ybar=4*\pgflinewidth,
        bar width=20.0pt,
        ylabel = {Accuracy},
        symbolic x coords={SVM,
        				      \kNN,
				      RF,
				      MLP},
	y tick label style={
    		/pgf/number format/.cd,
   		fixed,
   		fixed zerofill,
    		precision=0},
        xtick = data,
        x tick label style={
        		rotate=60,
		font=\footnotesize,
		anchor=north east,
		inner sep=0mm},
        nodes near coords,
        every node near coord/.append style={rotate=90, anchor=west, font=\footnotesize},
        enlarge x limits=0.2,
        legend cell align=left,
        legend style={
                at={(0.8775,0)},
                anchor=south,
                column sep=1ex
        }
    ]
\addplot[fill=red,opacity=1.00] 
coordinates { 
(SVM,88.88)
(\kNN,95.87)
(RF,98.23)
(MLP,93.97)
};
\addplot[fill=blue,opacity=1.00] 
coordinates { 
(SVM,51.90)
(\kNN,92.00)
(RF,92.87)
(MLP,80.24)
};
\legend{Level~1,Level~20}
\end{axis}
\end{tikzpicture}
\caption{A comparison of accuracy at level~1 and level~20}\label{fig:trend}
\end{figure}
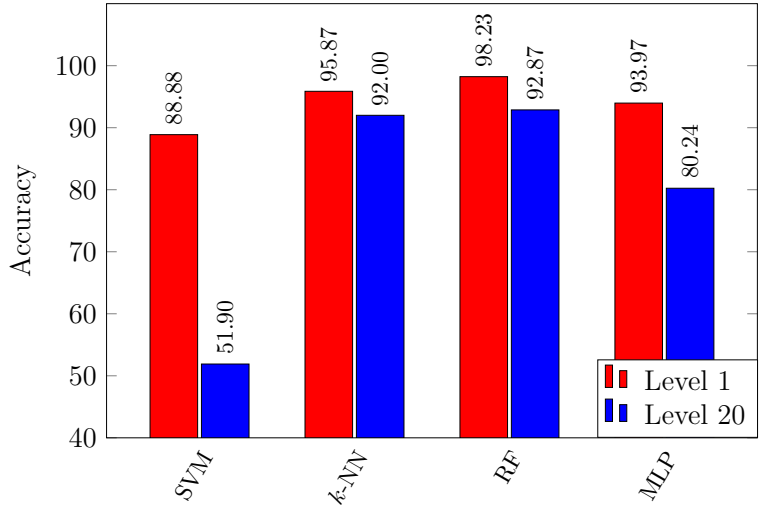

Figure~\ref{fig:trend} contains a bar graph comparing the performance of the most 
specific and the most general models for each of the four machine learning techniques considered,
based on bigram features.
As previously noted, SVM does not generalize well, even to level~2, and the
decline at level~20 shows that this only gets worse as the data becomes more general. 
This indicates that SVM is the least robust of the models considered,
at least for malware detection based on our byte bigram features.

MLP performed significantly better than SVM but nowhere
near as well as a random forest or \kNN. 
In fact, RF has the highest accuracy at every level,
which indicates that a random forest model (using~10 trees and
a maximum depth of~10) is the most robust technique considered. 
For both RF and \kNN, the decline between level~1 and level~20 is remarkably low.
It is interesting that for~\kNN, the level~20 case performs better than
the worst of the level~1 cases, whereas this is not true of RF,
where the range of model accuracies is generally much narrower.

Overall, we conclude that byte bigrams provide a strong feature for malware detection.
Moreover, when using neighborhood-based machine learning techniques
(e.g., \kNN\ and RF),
byte bigrams are a surprisingly robust feature, as evidenced by
their excellent performance in our multifamily experiments.

In this research, we considered~20 malware families with~1000 samples in each family. 
The dataset from which these samples were extracted is massive and is readily
available to other researchers. A wide variety of 
additional experiments could be considered.

We tested four classification techniques, namely, SVM, \kNN, random forest, and MLP. 
Each of these techniques---and especially MLP---include many parameters that could be further tuned.
Furthermore, many other machine learning techniques are applicable to this problem. 
For example, hidden Markov models (HMM), principal component analysis (PCA),
deep neural networks (DNN), and a wide variety of clustering techniques 
provide ample opportunity for additional useful experiments. 

Similar experiments to those conducted here could be conducted based on other features.
We have classified malware using bigram, 4-gram, and 6-gram features. 
Further research could include experimenting with other popular features, such
as opcodes, API calls, and so on to determine which features are most robust.
Combinations of features could also be considered.

Our experiments focused on binary classification of malware and benign samples
as the number of families in the malware class increased.
It would be interesting to see the results of
experiments that progressively combine, say, 20 adware families,
20 backdoor trojans, and so on. Perhaps a two-level detection strategy could be
developed, where we first classify by broad type (adware, backdoor, etc.), 
then classify into a specific family of that type. The results of such an approach could be compared
to more straightforward multiclass classification experiments. 

\bibliographystyle{plain}

\bibliography{references.bib}

\end{document}